\RequirePackage{fix-cm}
\documentclass[smallextended]{svjour3}       % onecolumn (second format)
\smartqed  % flush right qed marks, e.g. at end of proof
\usepackage{graphicx}
\usepackage{lineno,hyperref}
\usepackage{tabu}
\usepackage{graphicx}
\usepackage{algpseudocode}
\usepackage{balance}
\usepackage{subfigure}
\usepackage{booktabs}
\usepackage{mathtools}
\usepackage{amsmath}
\usepackage[linesnumbered, ruled,vlined]{algorithm2e}
\usepackage[export]{adjustbox}
\usepackage{amsmath}
\usepackage{mathabx}

\newcommand{\nop}[1]{}
\begin{document}

\title{Probabilistic Top-\textit{k} Dominating Queries in Distributed Uncertain Databases (Technical Report)
}


\author{Niranjan Rai         \and
        Xiang Lian %etc.
}


\institute{Niranjan Rai \at
              Kent State University \\
              \email{nrai@kent.edu}           %  \\
           \and
           Xiang Lian \at
             Kent State University\\
             \email{xlian@kent.edu}
}


\maketitle

\begin{abstract}
In many real-world applications such as business planning and sensor data monitoring, one important, yet challenging, task is to rank objects (e.g., products, documents, or spatial objects) based on their ranking scores and efficiently return those objects with the highest scores. In practice, due to the unreliability of data sources, many real-world objects often contain noises and are thus imprecise and uncertain. In this paper, we study the problem of \textit{probabilistic top-k dominating} (PTD) query on such large-scale uncertain data in a \textit{distributed} environment, which retrieves $k$ uncertain objects from \textit{distributed uncertain databases} (on multiple distributed servers), having the largest ranking scores with high confidences. In order to efficiently tackle the distributed PTD problem, we propose a MapReduce framework for processing distributed PTD queries over distributed uncertain databases. In this MapReduce framework, we design effective pruning strategies to filter out false alarms in the distributed setting, propose cost-model-based index distribution mechanisms over servers, and develop efficient distributed PTD query processing algorithms. Extensive experiments have demonstrated the efficiency and effectiveness of our proposed distributed PTD approaches on both real and synthetic data sets through various experimental settings.
\end{abstract}

\section{Introduction}
\label{sec:intro}

Nowadays, large volumes of imprecise and uncertain data have been produced in many real-world applications such as sensor networks, location-based services (LBS), moving object tracking, information extraction, and so on. The uncertainty in these application data is not uncommon, due to various reasons such as noises introduced during the data collection / transmission \cite{Cheng03}, inconsistent / conflicting public opinions/comments \cite{Arenas99,Fuxman05}, or the data perturbation/generalization \cite{LeFevre05} intentionally manipulated by humans for privacy-preserving purposes. Therefore, it is very important, yet challenging, to study how to manage and query such big uncertain data efficiently and effectively.

In this paper, we consider a useful and important spatial query, \textit{top-$k$ dominating query} \cite{yiu,xlian}, over a large-scale uncertain database in a distributed environment, which has many real-life applications such as multi-criteria decision making \cite{skoutas2009top}, coal mine surveillance \cite{xlian}, and so on. 

\begin{figure}[t]
\centering%\vspace{-2ex}
\subfigure[{\small certain data }]{
\scalebox{0.5}[0.5]{\includegraphics{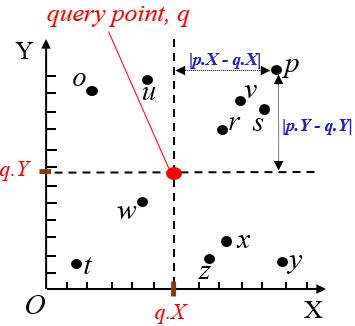}}
\label{subfig:certain}
}\qquad%
\subfigure[][{\small uncertain data }]{
\scalebox{0.5}[0.5]{\includegraphics{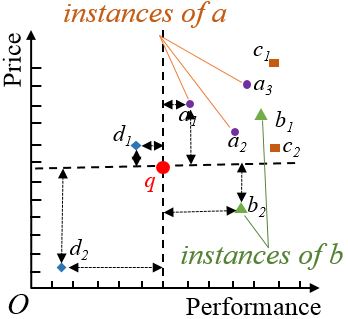}}
\label{subfig:uncertain}
}%\vspace{-2ex}
\caption{\small An example of top-$k$ dominating queries over certain and uncertain data.}
\vspace{-1ex}
\label{fig:cnu}
\end{figure}

Figure \ref{subfig:certain} shows an example of a top-$k$ dominating query over 2-dimensional (certain) data points $\{o, p, r, s, t, u, v, w, x, y, z\}$. Given a query point $q$, each object (e.g., $p$) has 2 \textit{dynamic attributes} w.r.t. $q$, given by $p.A_1 = |p.X  - q.X|$ and $p.A_2 = |p.Y  - q.Y|$. The top-$k$ dominating query retrieves $k$ objects with the highest ranking scores, where the score of each object is defined by the number of other objects that are worse than this object, in terms of dynamic attributes (i.e., the number of objects dynamically dominated by this object, as will be later discussed in Definition \ref{def:dynamic_dominance}). Intuitively, the query point $q$ is the preferred target specified by a user, and all objects are ranked based on the user's preferences w.r.t. $q$ (i.e., the ranking scores based on dynamic attributes).

%The top-$k$ dominating query obtains $k$ objects with the highest ranks (i.e., dynamically dominating the most objects). For different locations of the query point, the top-$k$ dominating result may also change. 

%if the value of k is 1, then top-$k$ dominating query returns the $w$ as the top-$1$ object w.r.t query point $q$. It is because in Figure \ref{subfig:certain}, object $w$ has the highest dominance score compared to other objects since it is closest to the query point $q$ and therefore has the highest score.

%{\color{Xiang}\bf add the explanation of top-k dominating definition here using a general figure in Fig. 1(a)..}

%Intuitively, in the top-$k$ dominating problem, the query point $q$ is the preferred target (e.g., ideal hotel) specified by users, and all other objects are ranked based on the users' preference (i.e., dynamic attributes w.r.t. query point). The top-$k$ dominating query obtains $k$ objects with the highest ranks (i.e., dynamically dominating the most objects). For different locations of the query point, the top-$k$ dominating result may also change. 

Due to the pervasive uncertainty in real-world data, in this paper, we will consider \textit{probabilistic top-$k$ dominating query} (PTD) \cite{xlian} over large-scale and distributed uncertain database \cite{cheng2004querying}, which obtains $k$ uncertain objects with the highest scores that are defined under the $possible ~worlds$ semantics \cite{dalvi2007efficient}. The PTD query has many real applications such as decision making in business. 

We have the following motivation example, shown as follows.

\begin{table}[ht]
\centering
\small
 \begin{tabu}{c | l | c}
 \hline
 {\bf Uncertain Object $t$} & \qquad {\bf Instance $t_j$} & {\bf Appearance Probability}  \\
 {\bf (Customer Group)} &  {\bf (Performance, Price)}  & {\bf $t_i.p$}\\
 \hline\hline
  & \qquad $a_1$ (7, 10) & 0.3\\
 $a$  & \qquad $a_2$ (9, 8) & 0.3\\
   & \qquad $a_3$ (10, 11) & 0.4 \\
 \hline
  $b$ & \qquad $b_1$ (8, 11) & 0.5 \\
    & \qquad $b_2$ (11, 9) & 0.5\\ 
 \hline
  $c$ & \qquad $c_1$ (11, 12) & 0.4\\
    & \qquad $c_2$ (11, 7) & 0.6 \\
\hline
  $d$ & \qquad $d_1$ (5, 8) & 0.2\\
    & \qquad $d_2$ (2, 2) & 0.8 \\
 \hline
\end{tabu}
\caption{\small The uncertain database in Figure \ref{subfig:uncertain}.}
\label{table1}%\vspace{-5ex}
\end{table}

%\vspace{0.5ex}
\noindent {\bf Example 1.} {\bf (Business Planning for the Laptop Market)} {\it Figure \ref{subfig:uncertain} shows an example of customers' preferences to laptops in a 2D performance-and-price space. As plotted in the figure, each customer group (e.g., a corporation or a school), say $a$, has a number of customers with different laptop preferences (e.g., $a_1$, $a_2$, and $a_3$), which can be obtained from user surveys. Table \ref{table1} depicts an uncertain database that contains customer preferences about laptops in Figure \ref{subfig:uncertain}, where customer groups correspond to uncertain objects, and laptop preferences by customers in each group can be treated as instances of the uncertain object (customer group). Each customer preference is associated with its price and performance attributes, as well as an appearance probability which indicates the portion of customers in the group who prefer this performance-and-price setting. As an example, instance $a_1$ from the customer group (uncertain object) $a$ has performance-and-price attributes (7, 10), which are preferred by 30\% of customers in the group $a$. 

Assume that a laptop company wants to produce a new laptop model, $q$, and would like to know which customer groups are more interested in this new model. As shown in Figure \ref{subfig:uncertain}, given a new laptop model (query point) $q$ (with targeted performance and price), the company can conduct a PTD query over the uncertain database in Table \ref{table1} to obtain $k$ customer groups that are expected to have preferences closer to the new model $q$ than most of other groups.}

There are many other applications for the PTD query, such as ranking $NO_2$ and $SO_2$ concentrations in the air \cite{park}, and the coal mine surveillance application \cite{xlian}. %{\color{Xiang} \bf Expand another example of PTD query}

Inspired by examples above, in this paper, we will study the PTD problem under a distributed setting. In particular, due to the large scale of uncertain databases or inherent features that data are collected and stored at different servers, we assume that the uncertain database is distributed on multiple servers. The distributed PTD problem retrieves $k$ uncertain objects with the highest (expected) ranking scores from distributed uncertain database (stored on different servers), using the MapReduce framework. Compared with centralized PTD query processing over a single server, PTD on the distributed uncertain database can achieve lower response time, by taking advantage of parallel computations among multiple servers. 

The distributed PTD problem is challenging, due to the large scale, uncertainty, and the distributing environment of the underlying data. One straightforward method is to simply collect all the uncertain objects from distributed servers, compute scores of all uncertain objects, and finally return $k$ uncertain objects with the highest ranking scores. However, this straightforward method incurs high communication cost which consumes much network bandwidth (due to the heavy workload of the data transfer among servers). Moreover, since the score calculation of each object involves all other objects, it is also rather costly (i.e., with quadratic cost) to compute exact scores for all uncertain objects, especially in large-scale distributed uncertain database. 

In order to efficiently and effectively tackle the PTD problem over the distributed uncertain database, in this paper, we will divide data into different partitions, create an index over each partition, and propose a novel cost-model-based index distribution technique to reduce the  search space of the PTD problem. We will also devise an efficient algorithm under a MapReduce framework that can effectively query on distributed uncertain databases and return the top-$k$ dominating uncertain objects.

In summary, we make the following contributions in this paper.%\%vspace{-1ex}
\begin{itemize}
\item We formally define the problem of probabilistic top-$k$ dominating query (PTD) over distributed uncertain database in Section \ref{sec:prob_def}.

\item We devise effective pruning methods for distributed PTD processing in Section \ref{sec:pruning}, and propose a MapReduce framework for answering PTD queries in the distributed uncertain database in Section \ref{sec:framework}.

\item We design an effective indexing mechanism for distributed PTD query processing and provide a cost model for guiding index distribution in Section \ref{sec:partitioning_indexing}.

\item We  propose efficient algorithms in Section \ref{sec:PTD_processing} for the filtering-mapper, filtering-reducer, and refinement-mapper phases which follow the MapReduce framework.

%\item We provide cost models for guiding index distribution for each server in Section \ref{sec:cost_model}. 
\item We evaluate the performance of our PTD approaches through extensive experiments in Section \ref{sec:exper}.
\end{itemize}%\%vspace{-1ex}

Section \ref{sec:related_work} reviews the related works on top-$k$ dominating queries on centralized certain/uncertain data and distributed certain data, as well as queries on distributed uncertain data. Section \ref{sec:conclusion} concludes this paper.

%\%vspace{2ex}
\section{Problem Definition}
\label{sec:prob_def}

In this section, we define \textit{probabilistic top-$k$ dominating query} (PTD) on the distributed uncertain database.

%\%vspace{2ex}
\subsection{Distributed Uncertain Database} 
An uncertain database consists of uncertain objects, each of which contains a number of \textit{mutually exclusive} instances associated with existence probabilities. Due to large scale of the uncertain database, we assume that uncertain objects reside in $N$ servers/nodes at different physical locations and connected by a network, which is called a distributed uncertain database.

\begin{definition}
\textbf{(Distributed Uncertain Database)} A distributed uncertain database $D$ is a collection of $N$ uncertain databases $D_1$, $D_2$,....., and $D_N$, which are located in $N$ servers in a network, respectively. Each uncertain database $D_l$ ($1\leq l\leq N$) contains $M_l$ uncertain objects. Each uncertain object $t \in D_l$ can be represented by an instance set $\{t_1,t_2,....,t_{|t|}\}$, and each instance $t_i$ ($1 \leq i \leq |t|$)  in $t$ contains $d$ numerical attributes $t_i.A_1$, $t_i.A_2$, ..., and $t_i.A_d$, as well as its appearance probability $t_i.p$, satisfying $\sum_{i=1}^{|t|} t_i.p \leq 1$.\label{def:distributed_uncertain_database}
\end{definition}

\begin{figure}[t!]
    \centering%\%vspace{-2ex}
    \includegraphics[width=280pt]{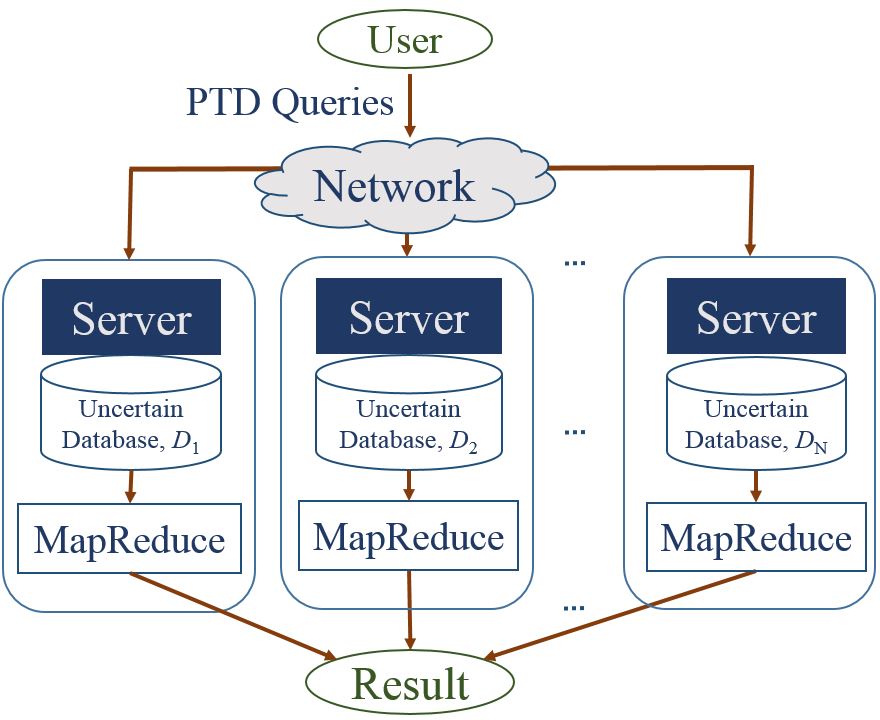}
    %\%vspace{-2ex}
    \caption{\small The PTD framework in the distributed uncertain database.}%\%vspace{-2ex}
    \label{fig:udb_dist}
\end{figure}

\noindent {\bf Example 2.} \textit{Figure \ref{fig:udb_dist} depicts an example of a distributed network that connects $N$ uncertain databases $D_i$ ($1\leq i\leq N$) located at $N$ server nodes, respectively. All these uncertain databases collectively form a distributed uncertain database $D$. Each uncertain database $D_i$ contains uncertain objects, for example, $a$, $b$, $c$, and $d$, as shown in Table \ref{table1}}. %$\qquad$  %$\square$

%\%vspace{2ex}
\subsection{Dynamic Dominance}
In this subsection, we define the dynamic dominance relationship between two instances in the distributed uncertain database.

\begin{definition}
{ \bf (Dynamic Dominance Between 2 Instances) } Given two instances $u_i$ and $v_j$ from uncertain objects $u$ and $v$, respectively, and a query point $q$, we say that $u_i$ \textit{dynamically dominates} $v_j$ with respect to $q$, denoted as $u_i \prec_q v_j$, if two conditions hold: (1) for every dimension $k$ ($ 1 \leq k \leq d $), we have $ |u_i.A_k - q.A_k| \leq |v_j.A_k - q.A_k| $ and (2) there exists at least one dimension $k$ ($1\leq k\leq d$) such that $ |u_i.A_k-q.A_k| < |v_j.A_k - q.A_k|$ holds.\label{def:dynamic_dominance}
\end{definition}

Intuitively, in Definition \ref{def:dynamic_dominance}, the query point $q$ is a preferred target of the user (e.g.,  targeted hotel or expected new product), based on which the dynamic attributes of other instances are calculated. We thus define the dynamic attributes of an instance, for example, $u_i$, as the absolute coordinate differences between $u_i$ and query point $q$ (i.e., $|u_i.A_k - q.A_k|$). If all dynamic attributes of instance $u_i$ are not worse than that of an instance $v_j$, and $u_i$ has at least one dynamic attribute strictly better than $v_j$, then we say that instance $u_i$ dynamically dominates instance $v_j$ with respect to $q$.

\vspace{1ex}
\noindent {\bf Example 3.} \textit{In Figure 3, the query point $q$ is the preference object used for defining dynamic attributes of instances. Instance $a_1$ dynamically dominates instance $c_1$, since $|a_1.A_k - q.A_k|$$<$$|c_1.A_k - q.A_k|$ holds for both dimensions $k= 1, 2$.} %$\square$

Equivalently, in Example 3, with respect to $q$, we can draw symmetric points $a_1'$, $a_1''$, and $a_1'''$ of instance $a_1$. Then, we can define the \textit{dynamic dominance regions}, DDR($a_1$), with respect to $a_1$, $a'_1$, $a''_1$, and $a'''_1$, which are shown as the shaded regions in Figure \ref{fig:dynamicdominance}. This way, any instance (e.g., $c_1$) falling in DDR($a_1$) is dynamically dominated by instance $a_1$.

%\%vspace{2ex}
\subsection{Ranking Scores of Instances/Objects }

In this subsection, we first provide the score definition for an instance of an uncertain object, and then formally define the ranking score of an uncertain object in an uncertain database. Since each uncertain object contains one or multiple instances with appearance probabilities, we need to take into account the probabilistic information when defining the score of an instance or an uncertain object.

\begin{definition}
\textbf{(The Score of an Instance)} For a given instance $t_j$ of an uncertain object $t$, its score, $S(t_j)$, is given by the appearance probability $t_j.p$ of $t_j$ times the summation of appearance probabilities for instances that are dynamically dominated by instance $t_j$:
\begin{eqnarray} 
S(t_j) =   t_j.p \cdot \sum_{\forall s_i \in D\land s\ne t} Pr\{t_j \prec_q s_i\}.%\vspace{-3ex}
\end{eqnarray}\label{def:instance_score}
\end{definition}

Intuitively, the score of an instance is the product of the probability that this instance exists and the summation of appearance probabilities for all instances that are worse than this instance. Thus, higher appearance probability, $t_j.p$, of an instance $t_j$ or more instances dominated by $t_j$ leads to higher score $S(t_j)$ of instance $t_j$.

\begin{figure}[t!]
    \centering%\%vspace{-2ex}
    \includegraphics[width=140pt]{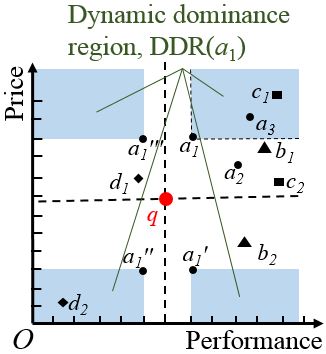}
    %\%vspace{-2ex}
    \caption{\small An example of dynamic dominance.}
    \label{fig:dynamicdominance}%\vspace{3ex}
\end{figure}

\vspace{1ex}
\noindent {\bf Example 4.} \textit{In Figure \ref{fig:dynamicdominance}, since instance $a_1$ dynamically dominates $c_1$ and $d_2$ w.r.t. $q$, by Definition \ref{def:instance_score}, the ranking score, $S(a_1)$, of instance $a_1$ is $a_1.p \cdot (d_2.p + c_1.p) = 0.3 \times (0.8 + 0.4) = 0.36$.} %$\quad$  $\square$}

After defining the score for the instance, we next give the score of the uncertain object by summing up the scores of its instances.

\begin{definition}
\textbf{(Ranking Score of an Uncertain Object)}  The ranking score $S(t)$ of an uncertain object $t$ is given by the summation of scores of all the instances $t_j$ in object \textit{t}, that is,

\begin{equation}
S(t) = \sum_{j=1}^{|t|} S(t_j).
\end{equation}
\label{eq:eq2}    
\label{def:object_score}%\vspace{-3ex}
\end{definition}

In Example 4, the score of instance $a_1$ is given by $S(a_1) = $ 0.36. Similarly, we have: $S(a_2)$ = 0.66 and $S(a_3)$ = 0.48. Thus, the score of uncertain object $a$ is given by $S(a)$ = $S(a_1)$ + $S(a_2)$ + $S(a_3)$ = 0.36 + 0.66 + 0.48 = 1.5.

%\%vspace{2ex}
\subsection{Probabilistic Top-$k$ Dominating Query in the Distributed Environment} 
Next, we define the \textit{probabilistic top-$k$ dominating query} (PTD) \cite{xlian} over a large-scale uncertain database in the distributed environment. Assume that we have $N$ servers where each server stores a subset, $D_i$ (for $1\leq i\leq N$), of the distributed uncertain database $D$.

\begin{definition}
\textbf{(PTD over the Distributed Uncertain Database)} Given a distributed uncertain database $D$ on $N$ server nodes, that is, $D$ = $\{D_1, D_2, ..., D_N\}$, and an integer $k$, a \emph{probabilistic top-$k$ dominating} (PTD) query retrieves $k$ uncertain objects, $t$, from the distributed uncertain database $D$ with the highest ranking scores $S(t)$, where $S(t)$ is given by Definition \ref{def:object_score}.\label{def:PTD}
\end{definition}

One straightforward method to tackle the PTD problem over the distributed uncertain database (as given by Definition \ref{def:PTD}) is to collect all uncertain objects from $N$ server nodes $D_i$ and return top-$k$ dominating uncertain objects over the distributed uncertain database $D$. However, in doing so, we must face several challenges. First, since uncertain data are distributed over different servers in the network, it is challenging to reduce the communication cost of the PTD query answering over the network. Second, due to the data uncertainty, efficient PTD query processing over distributed uncertain databases is also challenging, due to the exponential number of \textit{possible worlds} \cite{dalvi2007efficient}. In other words, it is not trivial how to efficiently retrieve the PTD answers over data associated with probabilities. Third, it is also challenging to design an effective and efficient distributed algorithm (e.g., following the MapReduce framework) to answer PTD queries over uncertain data.

%\%vspace{2ex}
\section{Pruning Methods for Distributed PTD Processing}
\label{sec:pruning}

In this section, we first provide the pruning heuristics for our distributed PTD query processing algorithms. Since the distributed uncertain database $D$ is located at $N$ different servers, in order to enable efficient PTD processing, we aim to filter out as many PTD false alarms as possible on each server locally, which results in only a few PTD candidates to be refined among $N$ servers (with minimal communication cost). These pruning methods can be also used for the filtering of false alarms over distributed servers.

%\vspace{1ex}
\noindent {\bf Distributed PTD Pruning.} Specifically, PTD over distributed uncertain database $D$ retrieves $k$ uncertain objects $t\in D$ with the highest scores $S(t)$ (as given in Eq.~(\ref{eq:eq2})). Thus, our basic idea is to retrieve at least $k$ PTD answer candidates from each partition $D_l$, and other objects in $D_l$ that are not candidates can be safely pruned.

\begin{lemma} (Distributed PTD Pruning) Given distributed uncertain database $D= \{D_1, D_2, ..., D_N\}$, if we conduct the PTD query over each partition $D_l$ on local server and obtain a set, $A_l$, of $k$ PTD answers (with the highest scores) only over $D_l$, then any uncertain object $t\in D_l$ that is not in $A_l$ can be safely pruned.
\label{lemma:lem1}%\%vspace{-2ex}
\end{lemma}
\noindent {\bf Proof.} {\it According to the lemma assumption, $A_l$ contains $k$ PTD answers over each local partition $D_l$. Then, the actual PTD answers ($k$ uncertain objects with the highest ranking scores) must be a subset of the retrieved $(N\cdot k)$ candidates from all the $N$ partitions. Therefore, other objects $t\in D_l$, which are not in the PTD answer set $A_l$, cannot be the actual PTD answers over $D$ and thus can be safely pruned.\qquad $\square$}

%\vspace{1ex}
\noindent {\bf Score Bound Pruning.} While the calculation of the score $S(t)$ for an object $t\in D_l$ may involve objects in $D$ located on $N$ different servers (in the worst case), it is challenging to efficiently compute the exact scores $S(t)$ across multiple servers with low communication cost. Therefore, alternatively, our basic pruning idea is to efficiently derive lower and upper bounds of the score $S(t)$, denoted as $t.LB$ and $t.UB$, respectively, and then utilize them to quickly rule out those uncertain objects $o$ with low ranking scores $S(o)$.

We have the following lemma about the pruning method by using the score bounds.
\begin{lemma} (Score Bound Pruning) Assume that we have obtained a set, $S_{cand}$, of $k$ PTD candidates $t^{(1)}$, $t^{(2)}$, ..., and $t^{(k)}$ in uncertain database $D_l \subseteq D$, and $\tau_l$ is the $k$ largest score lower bound among these candidates. Then, any uncertain object $o \in D_l$ can be safely pruned, if it holds that $o.UB < \tau_l$. 
\label{lemma:lem2}%\vspace{-2ex}
\end{lemma}
\noindent {\bf Proof.} {\it Since $o.UB$ is the upper bound of score $S(o)$, we have $S(o) \leq o.UB$. Similarly, we can obtain $t^{(i)}.LB \leq S(t^{(i)})$ for $k$ PTD candidates in $S_{cand}$, where $1\leq i\leq k$. 

Moreover, since $\tau_l$ is the $k$ largest score lower bound among $k$ candidates $t^{(1)}$, $t^{(2)}$, ..., and $t^{(k)}$, it holds that $t^{(i)}.LB \geq \tau_l$ for all objects $t^{(i)}$. By the transitivity, we can obtain $\tau_l\leq t^{(i)}.LB \leq S(t^{(i)})$ for all $1\leq i\leq k$. 

According to the lemma assumption that $o.UB < \tau_l$, by inequality transition, it holds that $S(o) \leq o.UB < \tau_l \leq S(t^{(i)})$ for $1\leq i\leq k$. That is, uncertain object $o$ has its ranking score $S(o)$ lower than at least $k$ PTD candidates in $S_{cand}$. Thus, based on Definition \ref{def:PTD}, object $o$ definitely cannot be in the PTD answer set, and can be safely pruned.  \qquad $\square$}

Lemma \ref{lemma:lem2} provides the pruning condition to quickly discard those uncertain objects with low ranking scores (i.e., $<\tau_l$). To enable this score bound pruning method, one important remaining issue is on how to obtain tight score lower/upper bounds (i.e., $t.LB$ and $t.UB$, respectively) of an uncertain object $t$ in a distributed environment. 

%\vspace{1ex}
\underline{{\it Derivation of Score Lower/Upper Bounds.}} Since the computation of the lower/upper bounds of score $S(t)$ involves objects across multiple servers, it is not possible to directly calculate the score, or its bounds, of an uncertain object $t$ over only one single partition $D_l$ at a local server. Therefore, as will be discussed later in Section \ref{subsec:index}, we will build an aggregate R-tree (aR-tree) index \cite{Lazaridis01}, $I_l$, over each partition $D_l$ and distribute it to other servers. This way, each server will store a partition $D_l$ (and its index $I_l$), and meanwhile maintain summaries (i.e., aR-tree) of $(N-1)$ other partitions on servers.

With partition $D_l$ and $N$ aR-tree indexes $I_1$, $I_2$, ..., and $I_N$, we next discuss how to compute the score lower/upper bounds, $t.LB$ and $t.UB$, of an uncertain object $t\in D_l$. In particular, since the score $S(t)$ is given by summing up the scores $S(t_j)$ of instances $t_j\in t$, the score lower/upper bounds $t.LB$ and $t.UB$ can be given by summing up score lower/upper bounds of instances, $t_j.LB$ and $t_j.UB$, respectively. That is, we have: 
\begin{eqnarray}
t.LB &=& \sum_{\forall t_j\in t} t_j.LB, \text{and}\label{eq:eq2.1}\\
t.UB &=& \sum_{\forall t_j\in t} t_j.UB.\label{eq:eq2.2}
\end{eqnarray}

Given a query point $q$, partition $D_l$, and $N$ indexes $I_1 \sim I_N$, the score lower/upper bounds, $t_j.LB$ and $t_j.UB$, of instance $t_j$ can be calculated as follows. With respect to $q$ and instance $t_j$, we can obtain the dynamic dominance region $DDR (t_j)$ (see an example in Figure \ref{fig:dynamicdominance}). We consider the bottom level of index $I_i$ that is stored on the current server. For any \textit{minimum bounding rectangle} (MBR) node $e$ in index $I_i$ (for $i\ne l$), if it holds that $e$ is fully contained in $DDR(t_j)$, then the number of uncertain objects, $e.sum$, under node $e$ will be counted in the score lower bound $t_j.LB$. Moreover, through index $I_l$ over partition $D_l$ (at local server), the summed probability of all instances in $DDR(t_j)$ will be added to $t_j.LB$. Formally, we have:
\begin{eqnarray}
t_j.LB &=& \sum_{\forall i\in [1, N] \wedge i\ne l} \sum_{\forall e\in I_i \wedge e \subseteq DDR(t_j)} e.sum\notag\\
&& +  \sum_{\forall o_r \in I_l \wedge o_r \subseteq DDR(t_j)} o_j.p.\label{eq:eq2.3}
\end{eqnarray}

Note that, in Eq.~(\ref{eq:eq2.3}), we underestimate the score $S(t_j)$ by ignoring those MBRs $e$ that are \textit{partially dominated} by $t_j$.  Here, given an instance $t_j$, an MBR $e$, and a query point $q$, we say that MBR $e$ is partially dominated by $t_j$ (with respect to query point $q$), if there exists at least one dimension $k$ $(1 \leq k \leq d )$, such that $ |e.A_k^- - q.A_k| \leq|t_j.A_k - q.A_k| \leq |e.A_k^+ - q.A_k| $ holds, where $e.A_k^-$ and $e.A_k^+$ are minimum and maximum possible values of MBR $e$ along the $k$-th dimension, respectively. Therefore, Eq.~(\ref{eq:eq2.3}) only considers those MBRs $e$ that are fully dominated by $t_j$, but not that partially dominated by $t_j$, which results in a score lower bound, $t_j.LB$.

Similarly, we can compute the score upper bound, $t_j.UB$, as the number, $e.sum$, of objects in those MBRs $e\in I_i$ intersecting with $DDR(t_j)$ (or the summed probability of instances that fall into $DDR(t_j)$). 
\begin{eqnarray}
t_j.UB  &=& \sum_{\forall i\in [1, N] \wedge i\ne l} \sum_{\forall e\in I_i \wedge e \bigcap DDR(t_j) \ne \emptyset} e.sum\notag\\
&& +  \sum_{\forall o_r \in I_l \wedge o_r \subseteq DDR(t_j)} o_j.p.
\label{eq:eq2.4}
\end{eqnarray}

Intuitively, Eq.~(\ref{eq:eq2.4}) computes a score upper bound $t_j.UB$, which overestimates the score, $S(t_j)$, by considering those MBRs $e$ both fully and partially dominated by $t_j$ with respect to query point $q$ (i.e., intersecting with $DDR(t_j)$).

%\vspace{2ex}
\section{The Distributed PTD Processing Framework}
\label{sec:framework}

In this section, we present a framework for PTD query answering over distributed servers in Algorithm \ref{alg:ptd_framework}. Specifically, our framework consists of four phases: \textit{offline pre-processing phase}, \textit{filtering-mapper phase}, \textit{filtering-reducer phase}, and \textit{refinement-mapper phase}.

In the first phase, we offline divide the uncertain database $D$ into $N$ disjoint partitions $\{D_1, D_2, ..., D_N\}$, based on our proposed cost model, and distribute them to $N$ servers, respectively (lines 1-2). For each partition $D_l$, we will construct an aR-tree index $I_l$, and send the index $I_l$ to $(N-1)$ other servers. In order to minimize the communication cost of the distributed PTD query processing (and save the space cost as well), we will also design a cost model to select the most appropriate levels of tree indexes $I_l$ and send to other servers (lines 3-5). 

In the filtering-mapper phase, given any PTD query, for each partition $D_i$, we will traverse the index $I_i$, and meanwhile apply the score bound pruning (discussed in Section \ref{sec:pruning}) to reduce the PTD search space and retrieve a candidate set $S_{cand}$ of PTD answers (line 7). Since each server containing partition $D_i$ only has indexes (summaries) of other partitions $D_l$ (for $l\ne i$), we cannot calculate the exact scores, $S(t)$, of candidates $t\in S_{cand}$. Thus, in this phase, we will send instances $t_j$ of candidates $t\in S_{cand}$, as well as those MBRs $eid$ in partition $D_l$ that are partially dominated by $t_j$, to the reducer server containing $D_l$ (line 8).

Next, in the filtering-reducer phase, each server containing partition $D_l$ will calculate the partial score of instances $t_j\in D_i$, with respect to the query point $q$ and MBRs $eid$ (line 10). In the refinement-mapper phase, each server will receive the partial scores of instances $t_j$ from the same uncertain objects $t$, and then combine partial scores into the exact scores, $S(t)$, of uncertain objects $t$ (lines 11-13).

Finally, $k$ uncertain objects $t$ with the highest scores $S(t)$ will be returned as PTD query answers (line 14).

%\vspace{-2ex}
\section{Offline Pre-Processing} 
\label{sec:partitioning_indexing}

In this section, we will discuss in Section \ref{subsec:index} how to perform the indexing in the offline pre-processing phase of Algorithm \ref{alg:ptd_framework}. Then, we will provide the cost model for guiding the index distribution among servers in Section \ref{subsec:cost_model}. 

%\vspace{2ex}
\subsection{Indexing Mechanism}
\label{subsec:index}

In this paper, we will randomly assign uncertain objects from the database to $N$ different partitions, denoted as $D_l$, which are then distributed to $N$ servers, respectively, where $1\leq l\leq N$. Note that, we tested other partitioning strategies (e.g., space partitioning or pivot-based partitioning), and found that the random partitioning can achieve the best pruning power, efficiency, and communication cost. 

%(experimental results are omitted due to space limitations)

\begin{algorithm}[t!]
\small
\KwIn{a distributed uncertain database $D$, the number, $N$, of servers, a query point $q$, and parameter $k$}
\KwOut{top-$k$ dominating uncertain objects with highest scores}
\tcp{\hspace{-1ex}Offline Pre-Processing Phase \hspace{-1ex}(Section \ref{sec:partitioning_indexing})}
$\{D_1, D_2, ..., D_N\}$  = {\sf Data\_Partitioning}($D$, $N$)\\
send each partition, $D_l$ ($1\leq l \leq N$), to a server\\
\For{each partition $D_l$}{
    construct an aR-Tree index, $I_l$\\
    send appropriate levels of index $I_l$ to ($N-1$) other servers
}
\tcp{Filtering-Mapper Phase (Section \ref{subsec:init})}
\For{each partition $D_i$}{
    apply pruning methods to retrieve a candidate set, $S_{cand}$, of PTD query answers from $D_i$ via index $I_i$\\
    send instances, $t_j$, of candidates $t\in S_{cand}$ to other partitions $D_l$ ($l\ne i$) that contain partially dominated MBRs $eid$
}
\tcp{Filtering-Reducer Phase (Section \ref{subsec:reducer1})}
\For{each partition $D_l$}{
    calculate the partial score of instances $t_j \in D_i$, w.r.t. query point $q$ and MBRs $eid$ 
}
\tcp{Refinement-Mapper Phase (Section \ref{subsec:mapper2})}
\For{each uncertain object $t$ without actual score}{
    receive partial scores of instances $t_j$ from the same uncertain objects $t$\\
    compute the actual score $S(t)$ of uncertain objects $t$
    }
    emit $k$ uncertain objects with the highest ranking score $S(t)$
%\vspace{-1ex}
\caption{The Distributed PTD Query Framework}\label{alg:ptd_framework} %\vspace{-1ex}
\end{algorithm}

To facilitate efficient PTD query processing on each server, we will build an \textit{aggregate R-Tree} (aR-Tree) \cite{Lazaridis01}, $I_l$, over each partition $D_l$. Specifically, for each uncertain object $t\in D_l$, we use a \textit{minimum bounding rectangle} (MBR) to minimally bound all its instances $t_j \in t$ (with existence probabilities $t_j.p \in[0, 1]$), and insert this MBR (associated with a COUNT aggregate: 1) into the aR-tree $I_l$. As a result, each leaf node $e$ in the aR-tree index contains MBRs of uncertain objects $t$ (each associated with a COUNT aggregate 1); each non-leaf node $e$ contains MBR entries $e_r$, which minimally bound all uncertain objects under $e_r$, and are associated with COUNT aggregates, $e_r.sum$, under entries $e_r$. 

\begin{figure}[t!]
\centering%\vspace{4ex} 
\subfigure[][{\small An example of MBRs }]{
\scalebox{0.5}[0.5]{\includegraphics{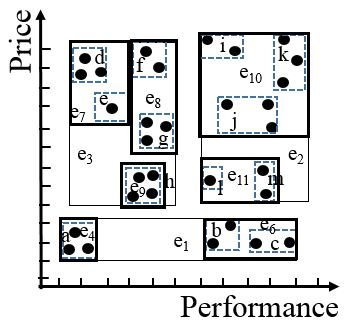}}
\label{subfig:arTree}
}\\
%\vspace{-2ex}\\
\subfigure[][{\small aR-Tree }]{\hspace{-3ex}
\scalebox{0.4}[0.4]{\includegraphics{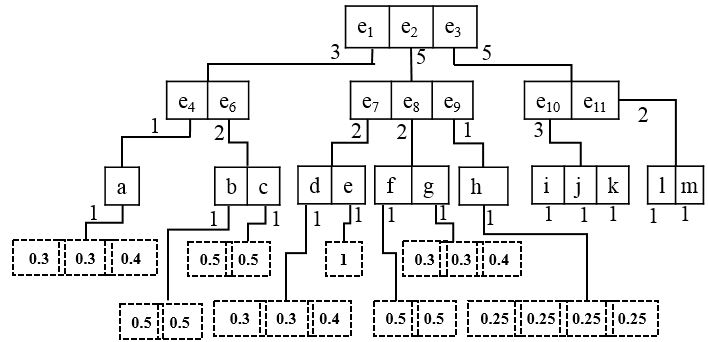}}
\label{subfig:arTree1}
}
%\vspace{-2ex}
\caption{\small The aR-Tree construction.}
%\vspace{2ex}
\label{fig:aR_Tree}
\end{figure}

Figure \ref{subfig:arTree} shows an example of MBRs that are built in a partition. In Figure \ref{subfig:arTree1}, we can see an example of aR-Tree built from the MBRs shown in Figure \ref{subfig:arTree}. Here, each uncertain object contains instances which has their existential probability (also known as appearance probability) associated with them. Thus, the aR-Tree nodes store the (SUM) aggregates of appearance probabilities of its child nodes. For example, node $e_4$ stores an aggregate 1, which is the summed appearance probability of all objects under $e_4$. Similarly, non-leaf node $e_6$ is associated with a SUM aggregate $2$.

%\%vspace{2ex}
\subsection{Cost-Model-Based Index Distribution}
\label{subsec:cost_model}

When the index $I_l$ over $D_l$ is constructed, we will send it to $(N-1)$ other servers (in order to calculate score lower/upper bounds for objects in other partitions). In order to minimize the communication cost, in this subsection, we will design a cost model to select the appropriate levels of the tree index $I_l$, that minimizes the \textit{communication cost}, $CC$, of PTD query answering, and distribute them to other servers. 

%In this subsection, we propose a cost model to estimate the \textit{communication cost}, $CC$, which can guide us to distribute indexes of appropriate levels to other servers with minimal $CC$ value.

Specifically, for each partition $D_l$, we build an aggregate R-Tree (aR-Tree) index and share this tree with other servers. While sending the index tree, we must decide whether to send the finer or coarser level of this tree to different servers. If we send the finer level of the tree, we will obtain tighter bounds of the scores, which leads to higher pruning power (i.e., smaller number of candidates). However, on the other hand, finer level results in more index nodes, which have higher chances to be partially dominated by instance and in turn higher communication costs. Similarly, if we send the coarser level of the index tree, then we may have smaller number of MBRs that might be partially dominated by the instance (i.e., incurring lower communication cost), but lower pruning power ($PP$) (i.e., leading to higher communication cost). So, we define a cost model which estimates the $CC$ for sharing different levels of the tree indexes and select the best levels of the index tree with minimal communication cost. Let $S_{MBR}(q)$ be a list of MBRs that cannot be dynamically dominated by other MBRs (w.r.t. query point $q$) in partition $D_l$. The Communication Cost, $CC$, of the distributed PTD processing can be given by:
\begin{equation}
 CC = \sum_{l=1}^N C_l, \label{eq:CC}
\end{equation}

\noindent where $C_l$ is the number of instances, $t_i$, of the uncertain objects, $t$, in a candidate set, $S_{cand}$, that partially dominate region of MBR $e \in S_{MBR}$. It can be calculated as:
\begin{eqnarray}
C_l = \sum_{\forall_{e \in S_{MBR}(q)}} N_{ins}(e,|S_{cand}|).\label{eq:instance_count}
\end{eqnarray}

In Eq.~(\ref{eq:instance_count}), function $N_{ins}(e, |S_{cand}|)$ returns the number of instances in the candidate set $S_{cand}$, which partially dominate the MBR $e$. We will collect the distribution of online query points from historical query log (note: if query logs are not available, we will use the Uniform distribution of query points to estimate the $C_l$ value), and take the average of the communication cost, $C_l$, over partition $D_l$, with respect to different possible locations of query points. $C_l$ in Eq.~(\ref{eq:instance_count}) plays a vital role in determining the communication cost of the PTD query, because higher value of $C_l$ leads to higher $CC$.

\begin{figure}[t!]
    \centering%\vspace{4ex}
    \includegraphics[width=160pt]{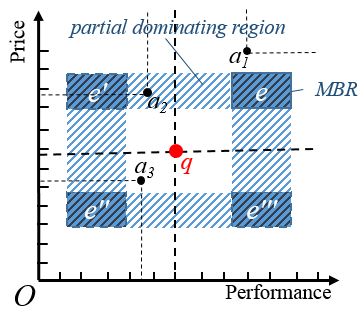}
    %\vspace{-2ex}
    \caption{\small An example of the partial dominating region (PDR).}
    \label{fig:partialdominance}%\vspace{2ex}
\end{figure}

Before we discuss how to compute function $N_{ins}(e, |S_{cand}|)$, we first introduce a notion, called \textit{partial dominating region} (PDR), with respect to a query point $q$ and an MBR $e$. Figure \ref{fig:partialdominance} illustrates an example of the PDR in the shaded region, in which $e'$, $e''$, and $e'''$ are symmetric mapping MBRs w.r.t. query point $q$ and MBR $e$. We say that the shaded region is a \textit{partial dominating region} (PDR), since any instance (e.g., $a_2$) falling into PDR must partially dominate MBR $e$ (and in turn incur the communication cost during PTD query processing). 

To calculate $N_{ins}(e, |S_{cand}|)$ for each MBR $e$, we will use a one-pass algorithm \cite{karp2003simple} to find those instances that partially dominate MBR $e$ (i.e., in the PDR), which should be output by the mapper phase. Here, we will consider the worst-case scenario, which scans all
the instances in PDR, and returns $|S_{cand}|$ uncertain objects with the highest numbers of instances falling in the PDR. This way, we can compute the total number of instances among the returned $|S_{cand}|$ objects that appear in PDR, which is exactly the output of function $N_{ins}(e, |S_{cand}|)$.

Finally, the only remaining issue is how to compute the number, $|S_{cand}|$, of candidate uncertain objects in each partition $D_l$, after the pruning. By considering the number of pruned objects for partition $D_l$ and applying \textit{central limit theorem} (CLT) \cite{wiki:clt}, we can obtain the number of the remaining candidates:
\begin{eqnarray}
    |S_{cand}|   &=&  |D_l| \cdot  (1- Pr\{t.UB < \tau_l ~|~ \forall{t \in D_l} \}) \notag\\
    &=& |D_l| \cdot  \left(1-\Phi\left( \frac{E(\tau_l)- \mu}{\sigma^2}\right)\right).\label{eq:scand}
\end{eqnarray}

\noindent where $\Phi(\cdot)$ is the \textit{cumulative density probability} (cdf) of standard normal distribution, $E(\tau_l)$ is the expected value of score threshold $\tau_l$ for partition $D_l$, and $\mu$ and $\sigma$ are the mean and variance, respectively, of a random variable for score upper bounds $t.UB$ of uncertain objects $t$ in $D_l$. %Due to space limitations, we omit the detailed derivations.

%\vspace{2ex}
\section{Distributed PTD Algorithms}
\label{sec:PTD_processing}
%\%vspace{-2ex}
The distributed PTD algorithm consists of MapReduce jobs in two rounds: the job in the first round includes Filtering-Mapper and the Filtering-Reducer functions, and the job in the second round is a map-only mapper function, namely Refinement-Mapper function. 

%\vspace{2ex}
\subsection{Filtering-Mapper Function}
\label{subsec:init}

In this subsection, we will present the algorithm for first Mapper function: Filtering-Mapper (as discussed in the PTD framework of Algorithm \ref{alg:ptd_framework}), which obtains at least $k$ PTD candidates over each partition $D_i$ on the $i$-th server (via indexes $I_l$ for $1\leq l\leq N$). 

%\%vspace{-2ex}

Specifically, Algorithm \ref{alg:init} illustrates the pseudo code of the filtering-mapper phase, which prunes false alarms of PTD answers at each uncertain database $D_i$, and obtains a set of at least $k$ PTD candidates from each server for refinement. In particular, the input of Algorithm \ref{alg:init} includes the partition number $i$ as key and uncertain database $D_i$, aR-Tree indexes $I_l$ of all other partitions, a query point $q$, and a parameter $k$ as the value, whereas the output of the algorithm includes a list of key-value pairs containing instances $t_j$ of PTD candidates $t$ and their partially dominated MBRs $eid$, which are also the input of the Mapper phase. 

\begin{algorithm}%[t!]
%\label{alg:mapper1}
\small
\KwIn{\textbf{Key}: partition number $i$, \textbf{Value}: uncertain database $D_i$, indexes $I_l$ of partitions $D_l$ ($1\leq l \leq N$), a query point $q$, and integer $k$ }
\KwOut{key-value pairs with PTD candidate instances and their partially dominated MBRs}
$S_{cand}$ = $\emptyset$, $cnt=0$, $\tau_i$ = $-\infty$;\\
initialize an empty min-heap $H$ with entry form $(e, key)$ \\
insert $(root(I_i), 0)$ into heap $H$\\
\While{$H$ is not empty}{
    $(e, key)$ = de-heap($H$)\\
    $e.UB= - key$;\\
    \If{$e.UB \leq \tau_i$} 
        {terminate the loop;}
    \eIf{$e$ is a leaf node}{
        \For{each uncertain object $t\in e$}{
            \eIf{$cnt<k$}{
                $t.LB$ = {\sf ScoreLB}($q, t, I_1, ..., I_N$) // Eq.(\ref{eq:eq2.1})\\
                $t.UB$ = {\sf ScoreUB}($q, t, I_1, ..., I_N$) // Eq.(\ref{eq:eq2.2})\\
                add ($t, t.LB, t.UB$) to $S_{cand}$;\\
                $cnt$++;\\
                \If{$cnt==k$} {let $\tau_i$ be the $k$-th largest score lower bound in $S_{cand}$}
            }
            {
                $t.UB$ = {\sf ScoreUB}($q, t, I_1, ..., I_N$) //Eq.(\ref{eq:eq2.2})\\
                                        \tcp{score bound pruning}
                \If {$t.UB > \tau_i$ }{
                    $t.LB$$=${\sf ScoreLB}($q, t, I_1, ..., I_N$)//Eq.(\ref{eq:eq2.1})\\
                    add ($t, t.LB, t.UB$) to $S_{cand}$;\\
                    update $\tau_i$ with the $k$-th largest score lower bound in 							$S_{cand}$\\
                    use $\tau_i$ to prune candidates $t' \in S_{cand}$ with 								$t'.UB<\tau_i$\\
                    $cnt = |S_{cand}|$;
                }
            }
        }
      
    }
    {   \tcp{$e$ is a non-leaf node}
        \For{each child node $e_r\in e$ }{ 
            $e_r.UB$ = {\sf ScoreUB}($q, e_r, I_1, ..., I_N$);\tcp{score bound pruning}
            \If{$e_r.UB > \tau_i$ }{
                insert ($e_r$, - $e_r.UB$) into heap $H$
            }
        }
        
    }
}
\For{ each uncertain object $t \in S_{cand}$}{
    \For{each instance $t_j\in t$}{
    obtain a list, $t_j.list$, of node/object IDs, $eid$, from indexes $I_l$ that are partially dominated by $t_j$, in the form ($l$, $eid$)\\
        \For{each entry $(l, eid)\in t_j.list$}{
            emit the key-value pair ($l$, \{$t$, $t_j$, $eid$, $t.LB$, $t.UB$, $\tau_i$\})
        }
    } 
}
\caption{Filtering-Mapper Function on Partition $D_i$}\label{alg:init}%\%vspace{-1ex}
\end{algorithm}

In Algorithm \ref{alg:init}, for each uncertain database $D_i$, we maintain an initially empty candidate set $S_{cand}$, the size, $cnt$, of the candidate set, and a pruning threshold $\tau_i$ (line 1). Moreover, we keep a (initially empty) \textit{minimum heap} $H$ with entries in the form $(e, key)$, where $e$ is an index node and $key$ is defined as the negative score upper bound, $e.UB$, of node $e$ (line 2). To retrieve PTD candidates, we first insert the root, $root(I_i)$, of the aR-tree index $I_i$ into heap $H$, in the form of $(root(I_i), 0)$ (line 3), and then utilize heap $H$ to traverse the index $I_i$ (lines 4-30). Each time we pop out an entry $(e, key)$ with the minimal key value from heap $H$ (line 5). If the score upper bound, $e.UB$, of the entry $e$ is less than or equal to threshold $\tau_i$, then we will terminate the loop (since all nodes in heap $H$ have scores lower than at least $k$ candidates; lines 6-8).

When node $e$ is a leaf node, for each uncertain object $t\in e$, we can compute its score lower/upper bounds, $t.LB$ and $t.UB$, and maintain a set, $S_{cand}$, of at least $k$ PTD candidates (lines 9-25). That is, we set $\tau_i$ to the $k$-th largest score lower bound in $S_{cand}$ (line 17 or 23), and use $\tau_i$ to filter out those objects $t$ with score upper bound not greater than $\tau_i$, by applying the score bound pruning method (lines 20-25). 

Similarly, when we encounter a non-leaf node $e$, we check each child entry $e_r\in e$ (lines 27-30). If it holds that  $e_r.UB \leq \tau_i$, then we can safely prune node $e_r$; otherwise, we insert entry ($e_r$, - $e_r.UB$) into heap $H$ for further checking (lines 29-30).

After the index traversal, we consider each candidate $t$ in the candidate set $S_{cand}$, and obtain a list of nodes $eid$ that are partially dominated by instances $t_j\in t$ (lines 31-33). If nodes $eid$ are from partition $D_l$, we will set the output key of this algorithm to $l$, and value to \{$t$, $t_j$, $eid$, $t.LB$, $t.UB$, $\tau_i$\}. We will emit the resulting key-value pairs to partition $D_l$ for further computation of scores of $S(t)$ (lines 34-35).

%\vspace{1ex}
\noindent {\bf Complexity Analysis.} The Filtering-Mapper function in Algorithm \ref{alg:init} traverses the index $I_i$ of partition $D_i$. Assuming that partition $D_i$ contains $M_i$ uncertain objects, the tree index $I_i$ has about $O\big(\frac{M_i}{F_i}\big)$ nodes, where $F_i$ is the average fanout in the R$^*$-tree $I_i$. Thus, in the worst case, the time complexity of traversing the index (lines 4-30) is given by $O(M_i)$ (by considering $F$ entries per node on average). Moreover, the time complexity of lines 31-35 is given by $O\big(|S_{cand}|\cdot \sum_{l=1}^N \frac{M_l}{F_l}\big)$, where $|S_{cand}|$ is the size of candidate set $S_{cand}$. Therefore, the total time complexity of Filtering-Mapper function is $O\big(M_i + |S_{cand}|\cdot \sum_{l=1}^N \frac{M_l}{F_l}\big)$.

The communication cost depends on the output of Algorithm \ref{alg:init}, which is given by $O(|S_{cand}| \cdot |t| \cdot |t_j.list|)$, where $|t|$ is the average number of instances per uncertain object, and $|t_j.list|$ is the average number of nodes/objects partially dominated by an instance $t_j$. 

%{\color{Niranjan} Each server $i$ has $M_i$ of uncertain objects. The run time complexity of Algorithm \ref{alg:init} is $O$($M_i\cdot log M_i$) for each server $i$.  Our PTD query uses the R*-Tree to index the uncertain objects so searching the tree takes $O(log M_i)$ time and the inner for loops runs for $O(M_i)$ time, hence, the run time is $O$($M_i\cdot log M_i$). The space time complexity is $O(M_i)$ since we store the information in R*-tree and store the candidate in candidate set, $|S_{cand}|$, both of them uses at max $O(M_i)$ space. The communication cost is $|S_{cand}| \cdot Size_{KV}$, where $Size_{KV}$ is the size of each $(key,value)$ pair emitted by Filtering-Mapper Function. }

\begin{algorithm}[t!]
\small
\KwIn{\textbf{Key}: partition ID $D_l$, \textbf{Value}: \{$t$, $t_i$, $eid$, $t.LB$, $t.UB$, $\tau_i$\}}
\KwOut{\textbf{Key}: object $t$~\textbf{Value}: \{$t.LB$, $t.\Delta S$, $\tau_i$\}}
    group all the instances $t_i$ based on their objectID $t$\\
    $t.\Delta S$ = 0 \\
    $\tau$ = $\max_{l=1}^N \tau_l$ \\
    \For {each instance $t_i \in t$}{
    $t_i.pScore$ = 0;\\
    load node $eid$ from index $I_l$\\
    \tcp{initialize heap $H$ with child nodes $e$ of $eid$}
    \For{each child $e \in eid$}{
        insert ($e$, $e.level$) into heap $H$
    }
    \While{H is not empty}
    {
    ($e, key$) = de-heap($H$)\\
    \eIf{$e$ is an uncertain object $o$}
    {
        \For{each instance $o_j \in o$ }{
            \If{$t_i$ dynamically dominates $o_j$}{
            $t_i.pScore$ = $t_i.pScore$ + $t_i.p \cdot o_j.p$;
            }
            }}
            {
            \For{each child $e_j \in e$}{
                \eIf{$t_i$ fully dominates $e_j$ w.r.t. $q$}{
                    $t_i.pScore$ = $t_i.pScore$ + $t_i.p \cdot e_j.sum$;
                }{
                    \If{$t_i$ partially dominates $e_j$ w.r.t. $q$}{
                        insert ($e_j$, $e_j.level$) into heap $H$
                    }
                }
            }
        }
        }
%        \tcp{calculate the partial score of instance $t_i$}\\
%        $t_i.pScore$ = $t_i.p$ $\cdot$ $t_i.pScore$\\ 
        \tcp{update the score upper bound of $t$}
         $t.UB$ = $t.UB$ - $eid.sum \cdot t_i.p$ + $t_i.pScore$\\
        \If{$t.UB < \tau$}
	    {emit ($t$, $\emptyset$) \\ 
	    return;
        }
        $t.\Delta S = t.\Delta S + t_i.pScore$
       }
emit the key-value pair ($t$, \{$t.LB$, $t.\Delta S$, $\tau$\})
\caption{Filtering-Reducer Function}\label{alg:reducer1}
\end{algorithm}

%\vspace{2ex}
\subsection{Filtering-Reducer Function}
\label{subsec:reducer1}

Algorithm \ref{alg:reducer1} shows the pseudo code of the reducer function: Filtering-Reducer (running on each of $N$ servers), which aims to compute the partial scores with respect to the received candidate instances $t_i$ and their partially dominated MBRs $eid$. It takes partition ID $D_l$ as the key and (object ID $t$, instance ID $t_i$, node ID $eid$, object score lower bound  $t.LB$, object score upper bound  $t.UB$) as the value. The output is key-value pairs (object ID $t$, \{object score lower bound $t.LB$, partial score $t.\Delta S$, threshold $\tau$\}).

In particular, the mapper algorithm first obtains all instances $t_i$ of the same PTD candidates $t$ from the filtering-mapper phase (line 1). We use $t.\Delta S$ to record partial scores w.r.t. instances $t_i$ and their partially dominated MBRs $eid$, which is initially 0 (line 2). Moreover, we let $\tau$ be the maximum value among all thresholds $\tau_l$ from input key-value pairs (i.e., $\max_{l=1}^N \tau_l$), which can be used to enable the score bound pruning (line 3). 

Next, we calculate the partial score $t.\Delta S$ for all the received instances $t_i \in t$ (lines 4-26). Initially, we load the partially dominated MBR node $eid$ from index $I_l$ (line 6), and insert children $e$ of node $eid$ into an initially empty min-heap $H$ in the form $(e, e.level)$ (lines 7-8), where $e.level$ is the level of the node $e$ (here, $e.level = 0$, when $e$ is a leaf node; $e.level > 0$, when $e$ is a non-leaf node). Intuitively, we set the key of the heap entry to $e.level$, since we want to quickly descend to leaf nodes during the index traversal. 

Each time we pop out an entry $(e, key)$ from the heap $H$ (lines 9-10). If $e$ is an uncertain object $o$, then we can calculate the partial score that $t_i$ dominates object $o$, which is given by summing up the existence probabilities of instances $o_j \in o$ dominated by $t_i$ (lines 11-14). On the other hand, if $e$ is a leaf or non-leaf node, then we check each child $e_j\in e$  (lines 16-21). If $t_i$ fully dominates $e_j$ (i.e., node $e_j$ is in $DDR(t_i)$), then we include aggregate $t_i.p\cdot e_j.sum$ in the partial score $t_i.pScore$ (lines 17-18); otherwise (i.e., $t_i$ partially dominates $e_j$), we insert $(e_j, e_j.level)$ into heap $H$ for further checking (lines 19-21). 

After computing the partial score w.r.t. $t_i$ and $eid$, we can now update and obtain tighter score upper bound, $t.UB$, of object $t$ (line 22). By applying the score bound pruning, if $t$ can be safely pruned (i.e., $t.UB<\tau$), then we emit $(t, \emptyset)$, which indicates that object $t$ can be ruled out (lines 23-25). 

Finally, for all instances $t_i$, we can obtain the total partial score $t.\Delta S$ (line 26), and emit a key-value pair in the form ($t$, \{$t.LB$, $t.\Delta S$, $\tau$\}) for combining partial scores of object $t$ in the next refinement round (i.e., Refinement-Mapper; line 27).

%\vspace{1ex}
\noindent {\bf Complexity Analysis.} The time complexity of Filtering-Reducer function in Algorithm \ref{alg:reducer1} is given by $O(|t|\cdot F_l^{eid.level})$, where $|t|$ is the number of instances in object $t$, $F_l$ is the average fanout of index $I_l$, and $eid.level$ is the level of node $eid$ in index $I_l$. 

The communication cost resulting from the Filtering-Reducer function is given by $O(1)$ (due to the emitted key-value pair in line 27).

\begin{algorithm}[t!]
\small
\KwIn{\textbf{Key}: $t$ ~\textbf{Value}: a list, $L$, of set values \{$t.LB, t.\Delta S, \tau$\}}
\KwOut{\textbf{Key}: $D_1$ ~\textbf{Value}: $\{t, S(t)\}$}
 \If{list $L$ contains an empty set $\emptyset$}{
    return;
 }

    $S(t)$ = $t.LB$ ;\\
    \For{each set \{$t.LB, t.\Delta S, \tau$\} $\in$ $L$}{
        $S(t)$ = $S(t)$ + $t.\Delta S$
    }
   \tcp{score bound pruning}
        \If{$S(t) < \tau$}{
            return;\\
        }
    emit($D_1$, $\{t, S(t)\}$)\\
\caption{Refinement-Mapper Function}\label{alg:mapper2}
\end{algorithm}

\subsection{Refinement-Mapper Function}
\label{subsec:mapper2}

Algorithm \ref{alg:mapper2} illustrates the pseudo code of a map-only job, Refinement-Mapper function, in a second round, which combines the partial scores, $t.\Delta S$, of instances and their partially dominated MBRs into an exact score $S(t)$. Specifically, the input of this algorithm takes object ID $t$ as the key, and a list, $L$, of set values ($t.LB$, $t.\Delta S$, $\tau$) as the value. Moreover, the output of this algorithm returns object $t$ and its exact score $S(t)$ to the master node.

In particular, for the received key-value pairs, if the list $L$ contains an empty set $\emptyset$ (which was emit by line 24 of Algorithm \ref{alg:reducer1}), it indicates that uncertain object $t$ can be safely pruned and thus ignored (lines 1-2). Otherwise, we first set $S(t)$ to the score lower bound $t.LB$ (line 3), and then add to $S(t)$ the partial scores $t.\Delta S$ for sets \{$t.LB, t.\Delta S, \tau$\} $\in$ $L$ (lines 4-5). If $S(t) < \tau$ holds, then we can apply the score bound pruning method (given in Lemma \ref{lemma:lem2}) to filter out uncertain object $t$; otherwise, we can emit object $t$ and its exact score $S(t)$ to the master server with $D_1$ (lines 6-8).

%\vspace{1ex}
\noindent {\bf Complexity Analysis.} The time complexity of the Refinement-Mapper function in Algorithm \ref{alg:mapper2} is given by $O(|t|)$, where $|t|$ is the number of instances in uncertain object $t$. Moreover, the communication cost (for line 8) is given by $O(1)$.

%\vspace{2ex}
\section{Experimental Evaluation}
\label{sec:exper}

In this section, we test our proposed distributed $PTD$ query framework over both real and synthetic data sets.

%\vspace{1ex}
\noindent {\bf Real/synthetic data sets.} We used real and synthetic datasets for our experimental evaluation. For real data, we tested California road networks \cite{tiger:data}, denoted as $Real$, which contains 98,451 MBRs for road segments in California. These MBRs can be considered as uncertainty regions of uncertain objects.

For synthetic data, we generate uncertain objects $t$ as follows. In particular, for each uncertain object $t$, we generate the center $C_t$ of its uncertainty region $UR(t)$ (bounding instances), following Uniform, Gaussian, or Zipf (with skewness 0.8) distribution. Then, we also randomly produce the side length of $UR(t)$ within $(0, l_{max}]$. This way, we can obtain uncertainty regions $UR(t)$ of uncertain objects $t$, and uniformly generate instances within $UR(t)$, where the number of instances per object is within $[|t|_{min}, |t|_{max}]$. With different data distributions, we can obtain 3 types of data sets, denoted as $uniform$, $gaussian$, and $zipf$, respectively. For each real/synthetic data set, we randomly distribute uncertain objects to $N$ servers and construct indexes over $N$ partitions.

%\vspace{1ex}
\noindent {\bf Measures.} To evaluate the PTD query performance, we will randomly generate 20 query points in the data space. We report the \textit{wall clock time} and \textit{communication cost}. Here, the wall clock time is the time cost of answering the PTD query by taking the average over 20 queries, whereas the communication cost is the average size of communication data among servers.

%\vspace{1ex}
\noindent {\bf Competitor.} We will compare our distributed $PTD$ algorithm with a baseline algorithm (mentioned in Section \ref{sec:intro}), which retrieves all data and indexes from other $(N-1)$ servers, and conducts PTD queries on a single server. 

%\vspace{1ex}
\noindent {\bf Parameter settings.} Table \ref{table:parameter} depicts the parameter settings, where the default values are in bold. Each time we vary the values of one parameter, while other parameters are set to their default values. We ran experiments on 10 dell PowerEdge R730 servers with Intel Xeon E5-2630 2.2GHz CPU and 64 GB memory. All algorithms were compiled by Javac 1.8 and used Hadoop 1.2.1 for MapReduce.

%The parameters are $k$, No. of servers (N), $l_{max}$, and size of dataset $(|D|)$. The different value of parameters are shown in the.

\begin{table}[t!]
\centering
\small
 \begin{tabu}{| c | l |}
 \hline
 {\bf Parameters} & \qquad\qquad {\bf Values}  \\
 \hline\hline
  $N$ & 1, 2, 5, 8, \textbf{10}\\
 \hline
  $l_{max}$ &  1, 2, \textbf{3}, 4, 5\\
 \hline
 $k$  & 5, 10, \textbf{15}, 20, 25\\
 \hline 
 $[|t|_{min}, |t|_{max}]$ & [2,5], [2,8], \textbf{[2,10]}, [2,12], [2,15]\\
\hline 
  $|D|$ & 100K, 200K, \textbf{500K}, 800K, 1M\\
\hline
\end{tabu}
\caption{\small The parameter settings.}\label{table:parameter}%\vspace{2ex}
\end{table}

%\vspace{2ex}
\subsection{The Cost Model Verification}

%{\color{Xiang} \bf  Niranjan, please add the cost model verification here!}

\begin{figure}[t!]
\centering%\vspace{-2ex}
\subfigure[][{\small communication cost (CC)}]{\hspace{-4ex}
\scalebox{0.065}[0.065]{\includegraphics{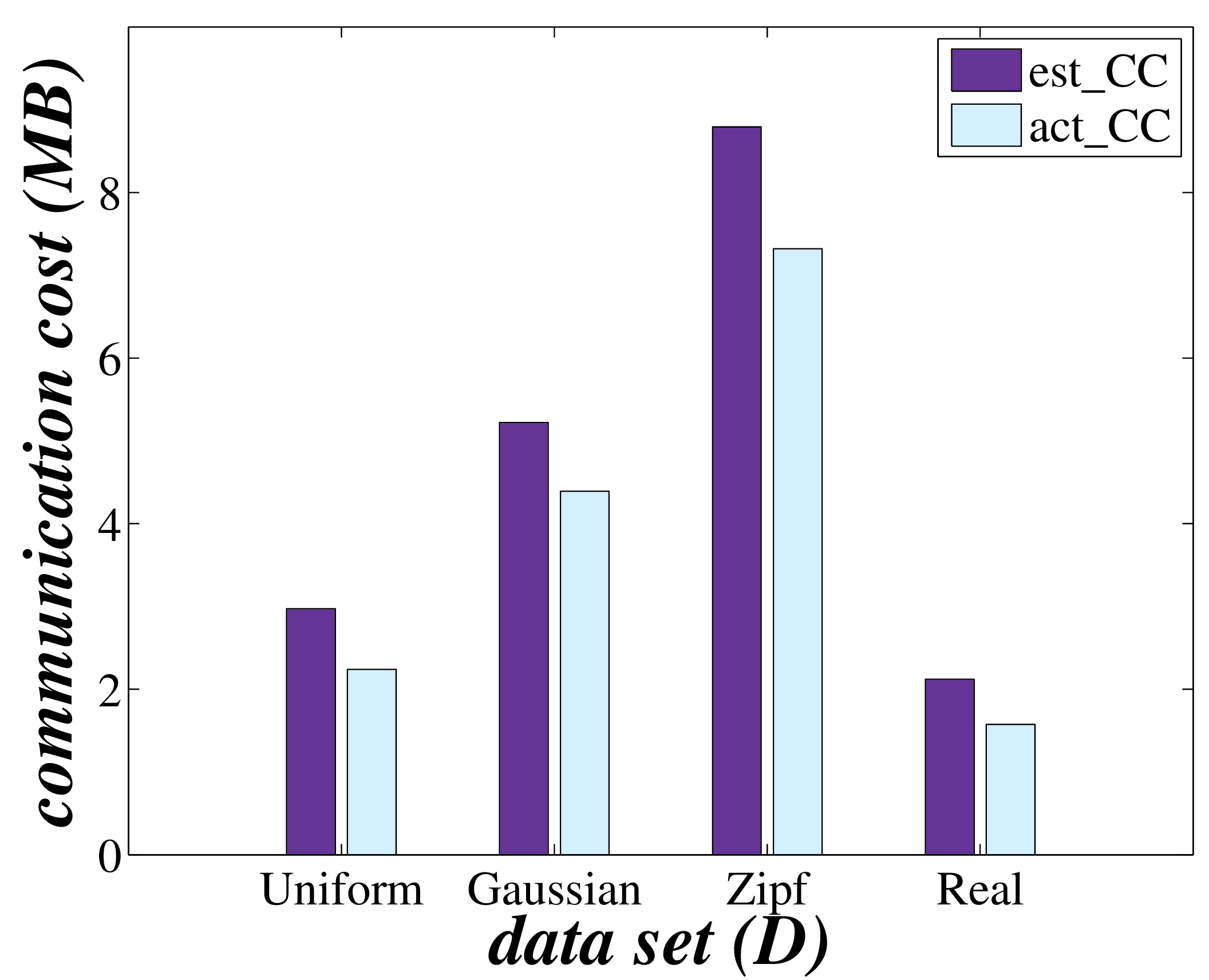}}\label{subfig:commcost}
}\qquad%
\subfigure[][{\small pruning power (PP)}]{
\scalebox{0.065}[0.065]{\includegraphics{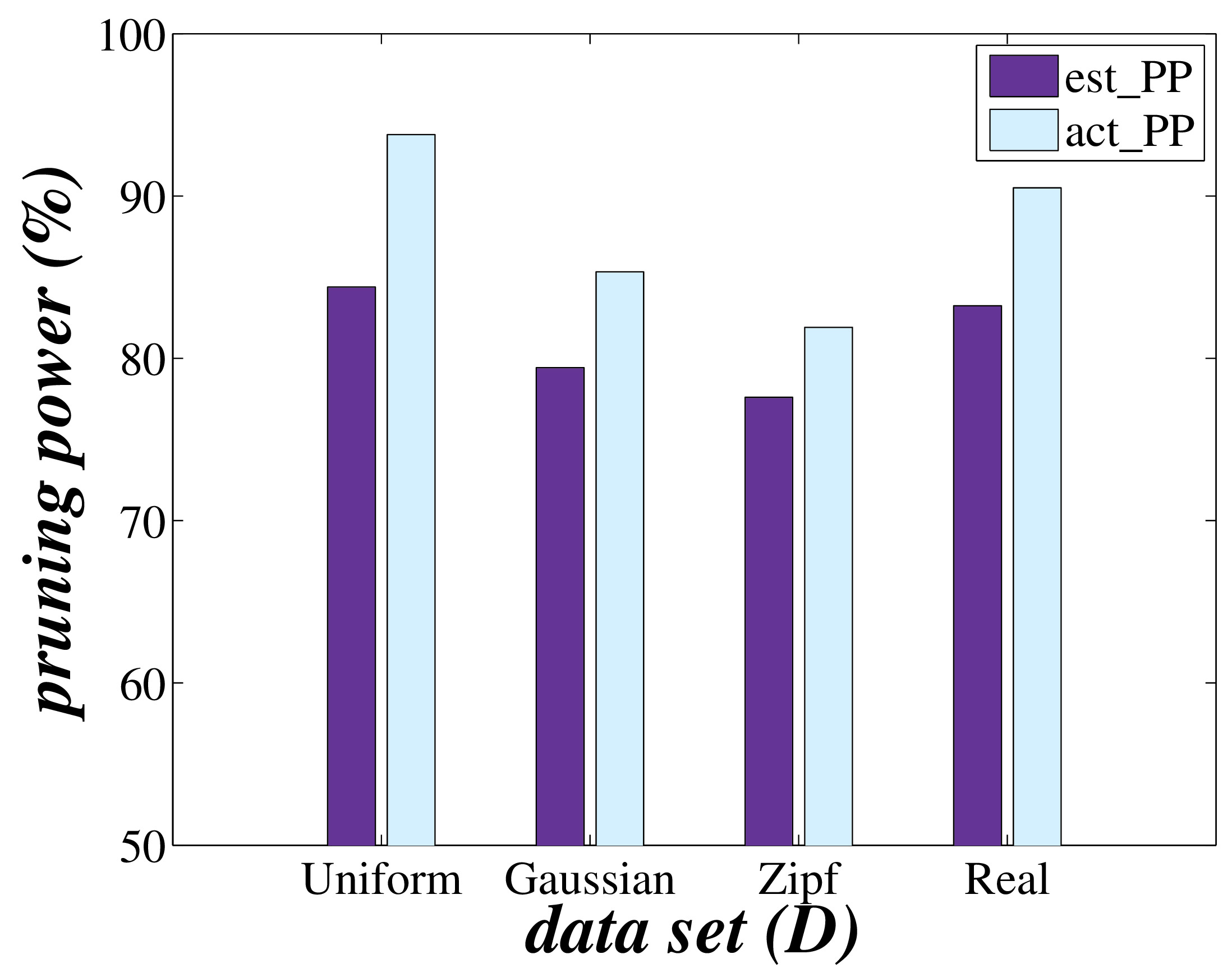}}\label{subfig:pruningpow}
}%\vspace{-2ex}
\caption{\small The cost model verification for the communication cost (CC) and pruning power (PP) vs. real/synthetic data sets.}%\vspace{-2ex}
\label{fig:costmodelveri}
\end{figure}

In the first set of experiments, we verify the effectiveness of our proposed cost model for the index distribution (as discussed in Section \ref{subsec:cost_model}). Specifically, Figure \ref{subfig:commcost} compares the estimated communication cost, denoted as $est\_CC$ (given by Eq.~(\ref{eq:CC})) with the actual one, $act\_CC$; Figure \ref{subfig:pruningpow} shows the comparison of estimated pruning power, $est\_PP$, with the actual one, $act\_PP$. Here, we obtain the estimated pruning power given by $est\_PP = \frac{|D_l|-|S_{cand}|}{|D_l|}$ for each partition $D_l$, where $|S_{cand}|$ is estimated by Eq.~(\ref{eq:scand}). From figures, we can see that both estimated communication costs and pruning powers over real/synthetic data sets can closely mimic the actual ones (and their trends), which confirms the correctness and effectiveness of our cost model (i.e., Eqs.~(\ref{eq:CC}) and (\ref{eq:scand})) for guiding the selection of levels in indexes. 

%Figure \ref{fig:costmodelveri} compares the result of our cost model. In  we show the comparison between estimated and actual pruning power of our PTD algorithm i.e. $est\_PP$ and $act\_PP$, whereas Figure \ref{subfig:commcost} shows the comparison of estimated communication cost $est\_CC$ and actual communication cost $act\_CC$. As you can see our PTD algorithm performs better than our estimation. }

%\%vspace{2ex}
\subsection{The Evaluation of the Efficiency}

\begin{figure}[t!]
\centering %\vspace{-2ex}
\subfigure[][{\small wall clock time}]{\hspace{-4ex}
\scalebox{0.065}[0.065]{\includegraphics{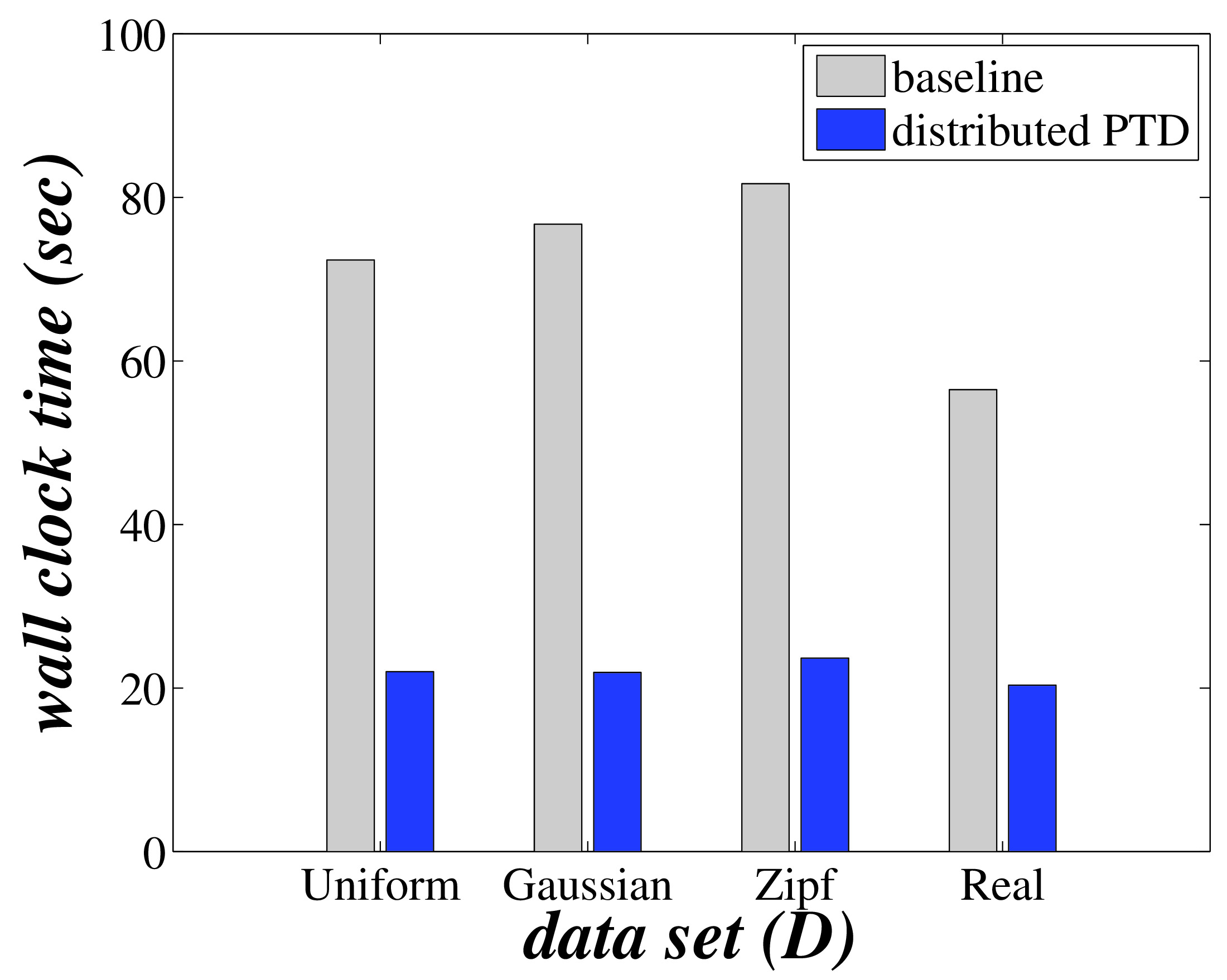}}\label{subfig:datasettime}
}\qquad%
\subfigure[][{\small communication cost}]{
\scalebox{0.065}[0.065]{\includegraphics{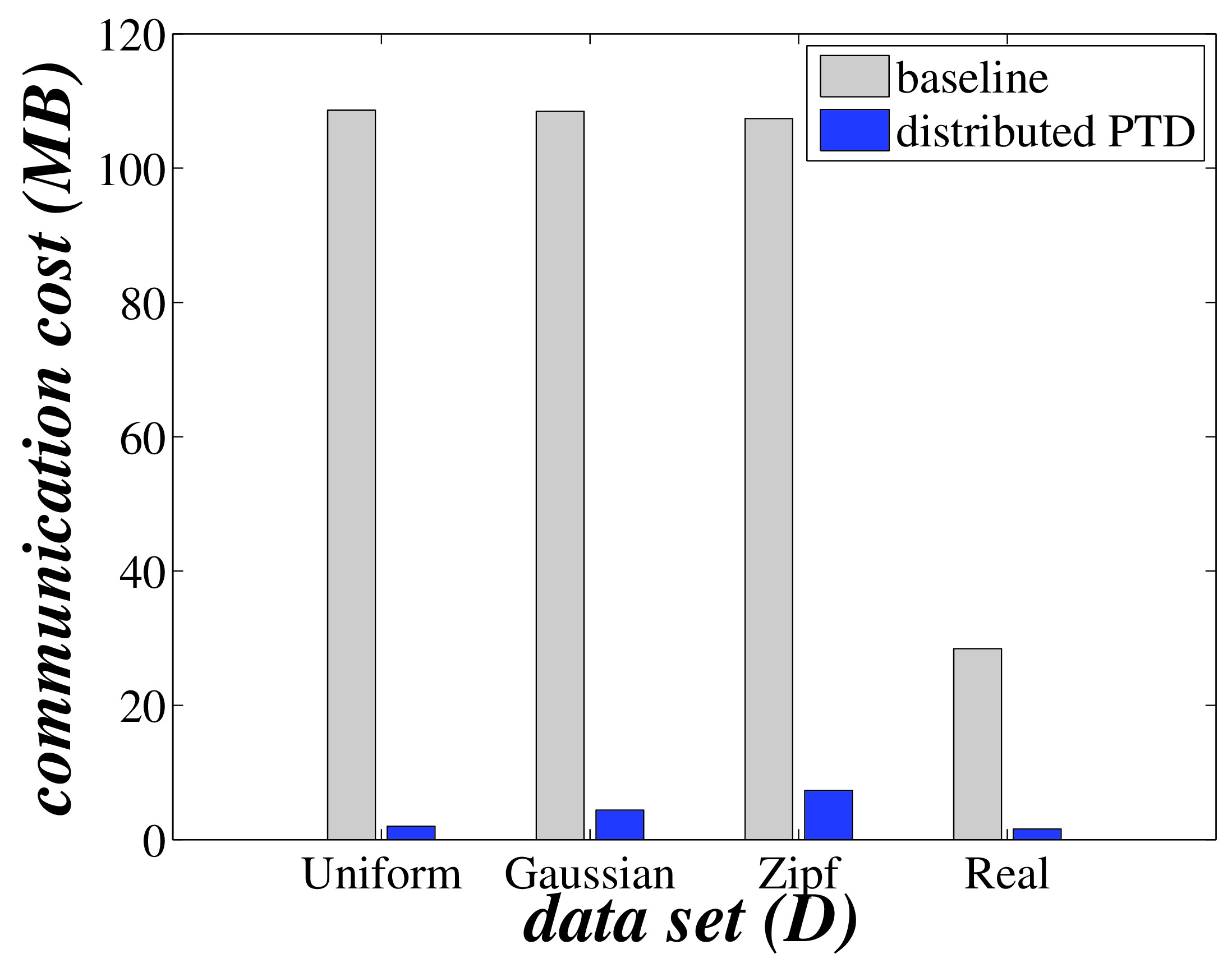}}\label{subfig:datasetcost}
}%\vspace{-2ex}
\caption{\small The PTD performance vs. real/synthetic data sets.}%\vspace{-2ex}
\label{fig:ptdvsbaseline}
\end{figure}

%\vspace{1ex}
\noindent {\bf The PTD performance vs. real/synthetic data sets.} Figure \ref{fig:ptdvsbaseline} shows the performance of our distributed $PTD$ algorithm and the baseline over real and synthetic data sets, where parameters are set to default values for synthetic data. From the figure, we can see that our $PTD$ algorithm outperforms the baseline algorithm by about an order of magnitude, in terms of both wall clock time and communication cost. This confirms the efficiency of our distributed $PTD$ approach over both real and synthetic data sets.

Below, to clearly show the trends and robustness of our $PTD$ approach with respect to different parameters, we will only report $PTD$ experimental results over synthetic data sets, by testing different parameters (such as $N$, $l_{max}$, $k$, $[|t|_{min}, |t|_{max}]$, and $|D|$). 

\begin{figure}[t!]
\centering%\vspace{-2ex}
\subfigure[][{\small wall clock time}]{
\scalebox{0.065}[0.072]{\includegraphics{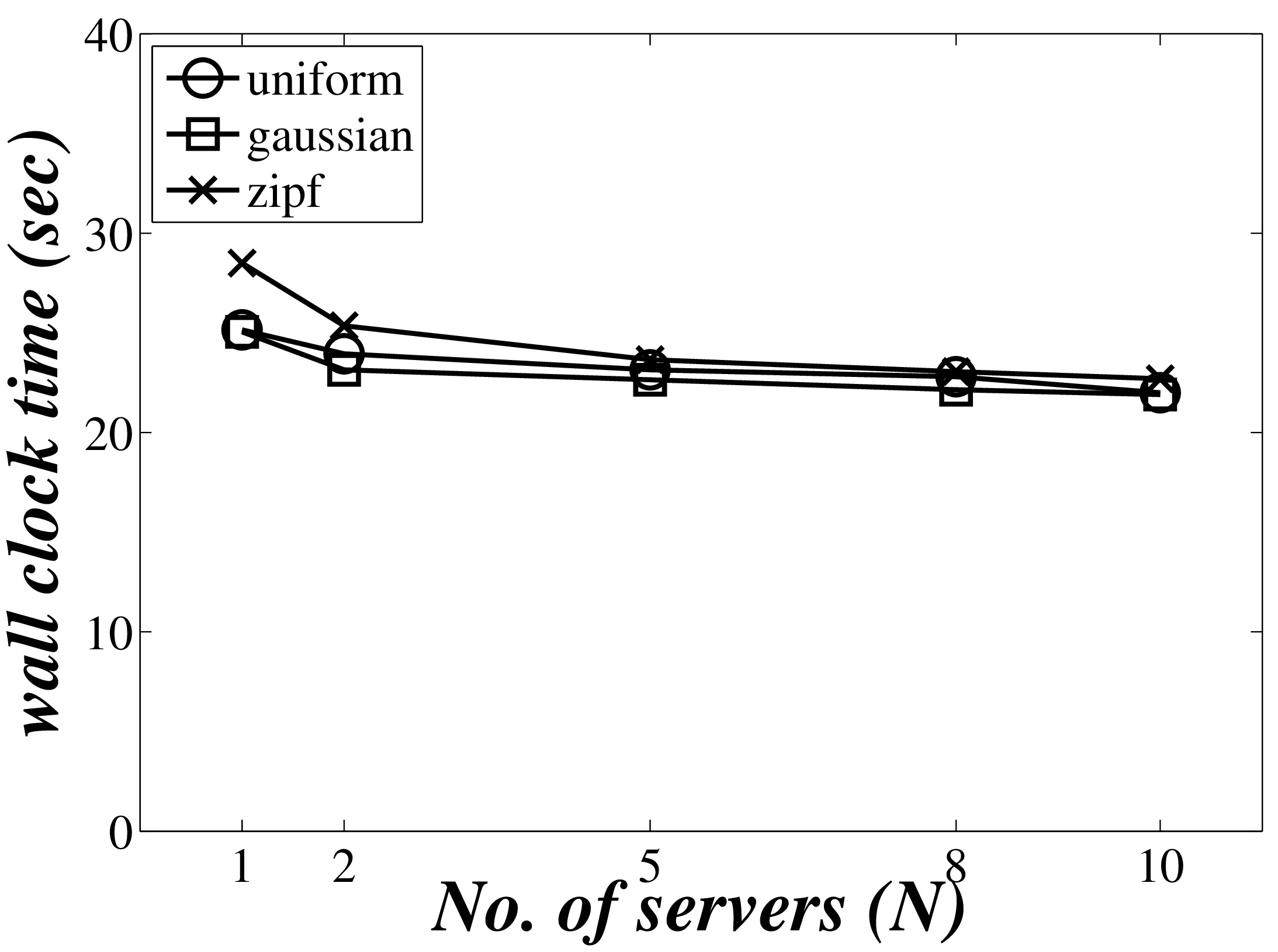}}\label{subfig:ntime}
}\qquad%
\subfigure[][{\small communication cost}]{
\scalebox{0.065}[0.065]{\includegraphics{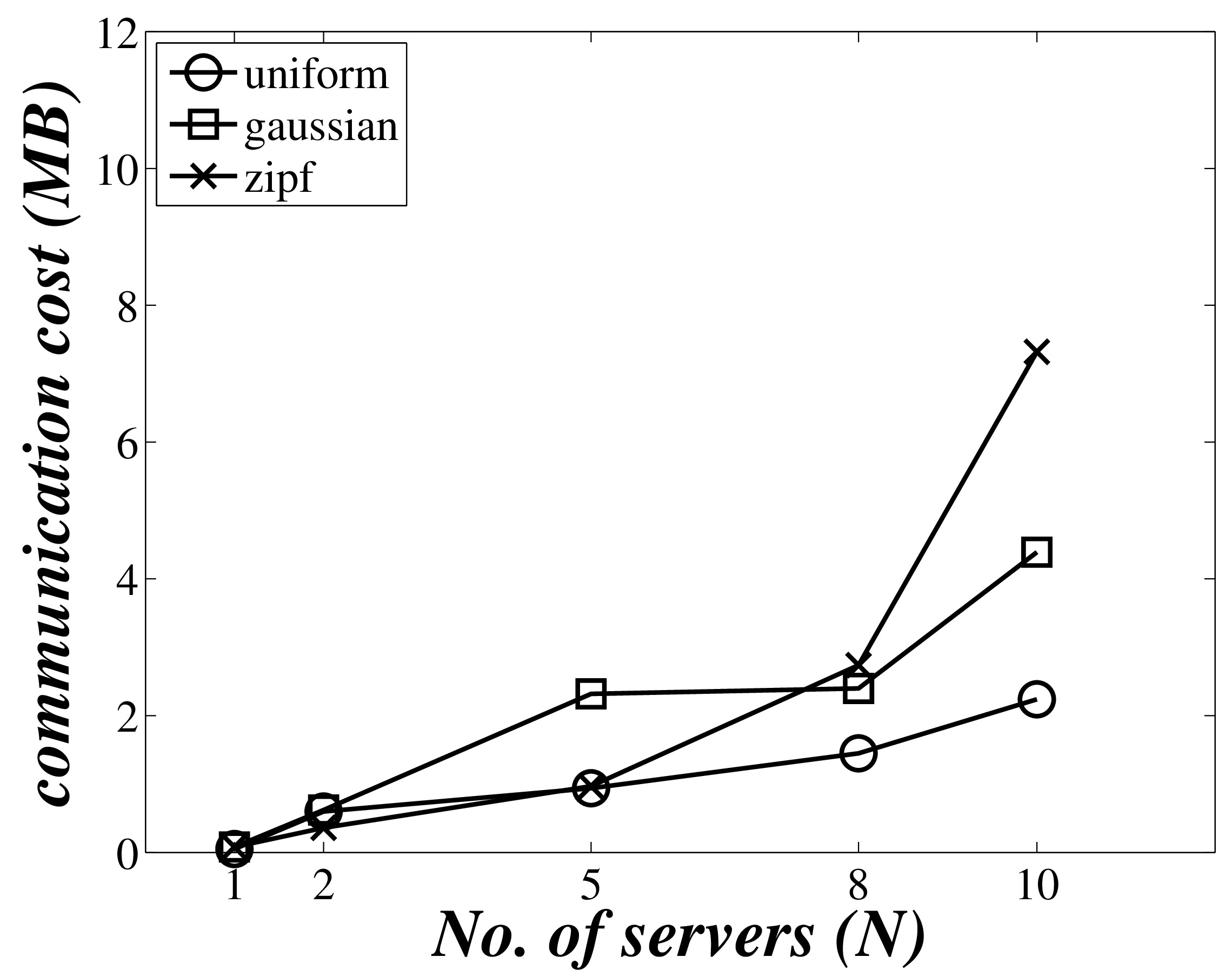}}\label{subfig:ncost}
}%\vspace{-2ex}
\caption{\small The PTD performance vs. No. of servers $N$.}%\vspace{-2ex}
\label{fig:ptdserver}%\vspace{2ex}
\end{figure}

%\vspace{1ex}
\noindent {\bf The PTD performance vs. No. of servers $N$.} Figure \ref{fig:ptdserver} evaluates the wall clock time and communication cost of our $PTD$ algorithm over $uniform$, $gaussian$, and $zipf$ data sets, by varying the number of servers (i.e., $N$) from 1 to 10, where default values are used for other parameters (as depicted in Table \ref{table:parameter}). From the figure, we can see that, running the $PTD$ query on a single server (i.e., $N=1$) is more costly. This is because the processing on one server cannot take advantage of parallel computation over multiple servers. With the increase of $N$, the amount of partition data in $D_l$ to be processed for each server decreases, and multiple servers can answer the $PTD$ query in parallel. Therefore, the wall clock time becomes smaller for larger $N$ value. On the other hand, since each server returns at least $k$ $PTD$ candidates, for more servers (i.e., larger $N$), the total number of $PTD$ candidates tends to increase, which leads to higher communication cost.

\begin{figure}[t!]
\centering%\vspace{-2ex}
\subfigure[][{\small wall clock time}]{\hspace{-4ex}
\scalebox{0.07}[0.07]{\includegraphics{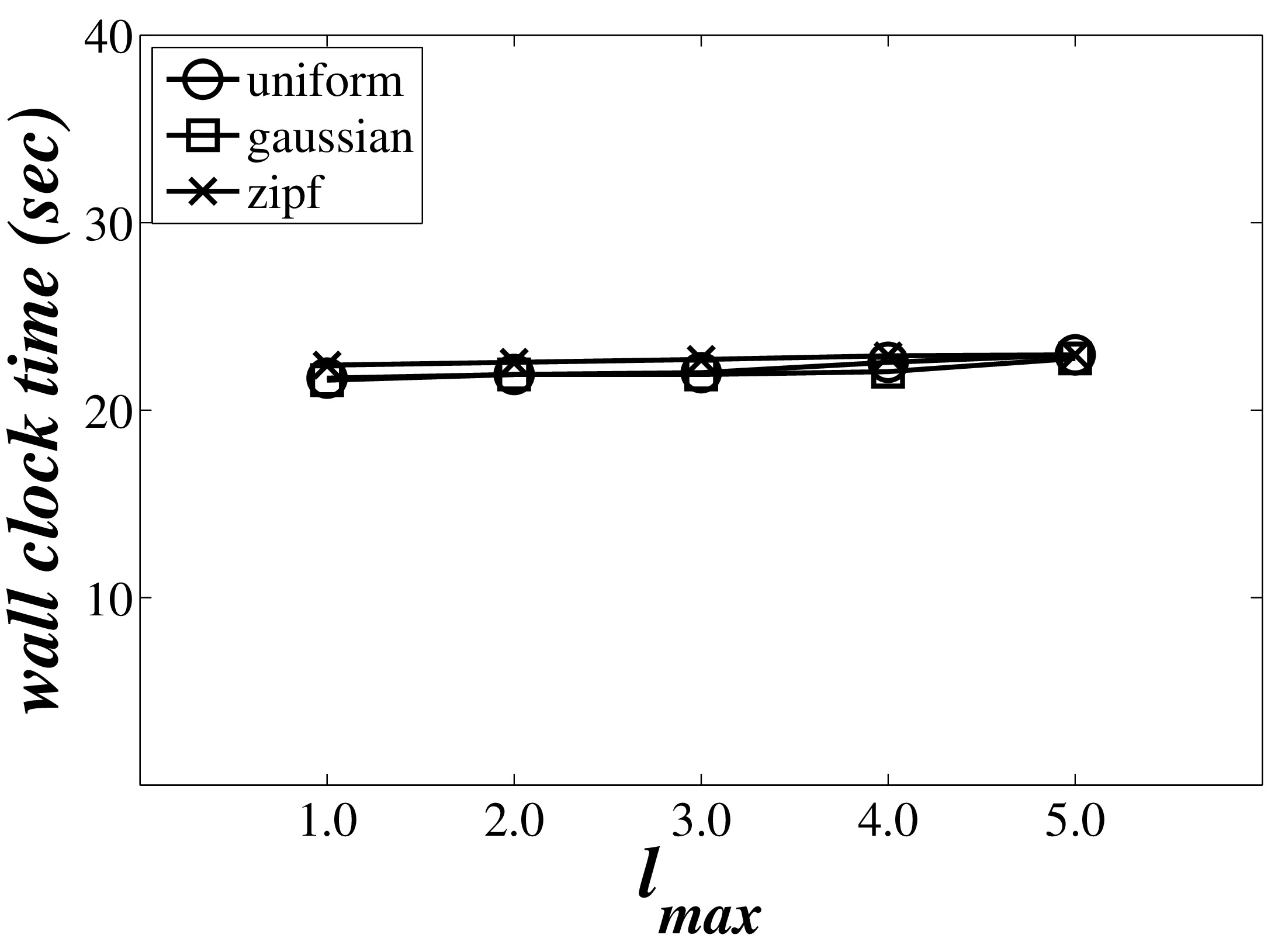}}\label{subfig:ltime}
}\qquad%
\subfigure[][{\small communication cost}]{
\scalebox{0.07}[0.07]{\includegraphics{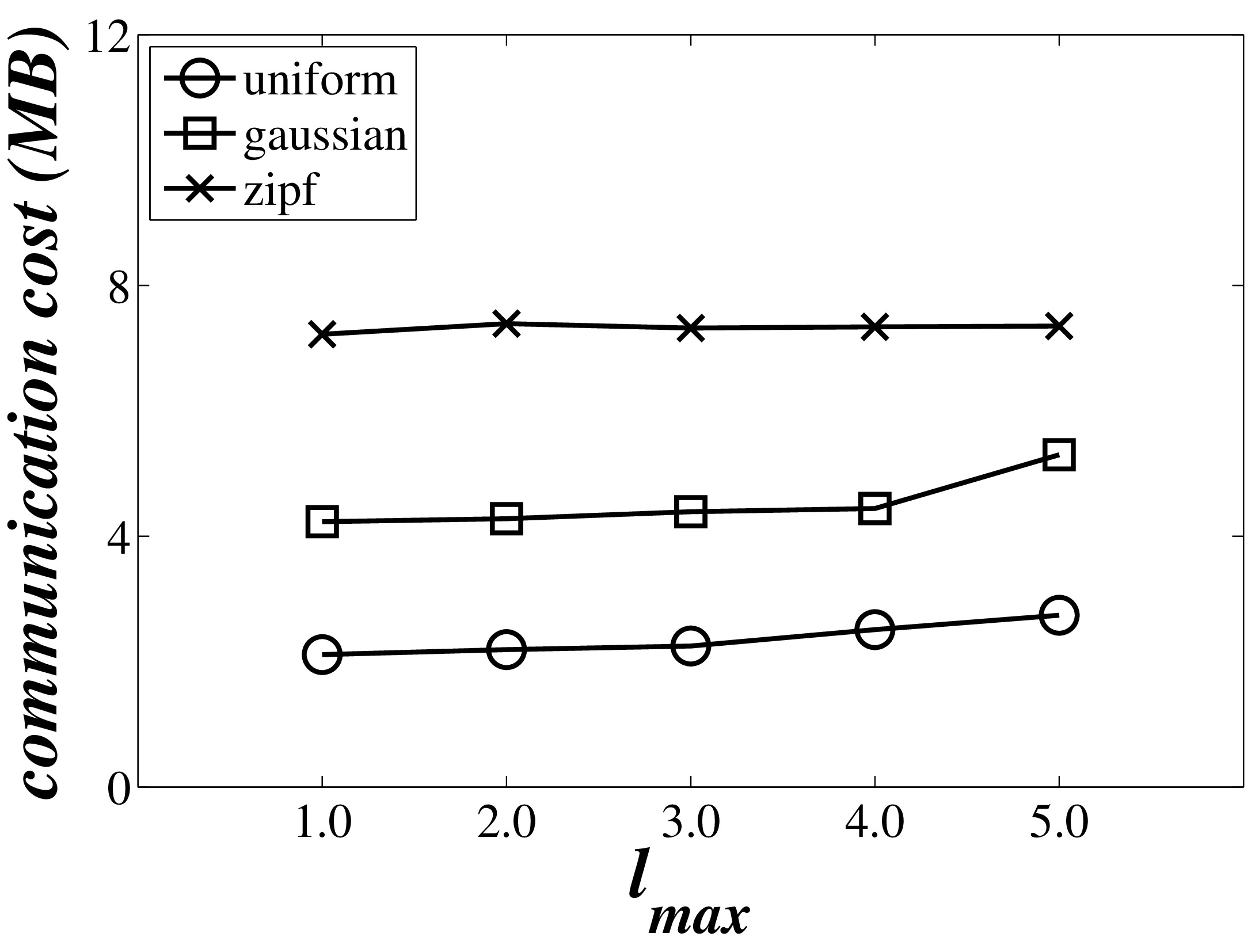}}\label{subfig:lcost}
}%\vspace{-2ex}
\caption{\small The PTD performance vs. $l_{max}$.}%\vspace{-2ex}
\label{fig:ptdlmax}%\vspace{2ex}
\end{figure}

%\vspace{1ex}
\noindent {\bf The PTD performance vs. maximum side length, $l_{max}$, of uncertain objects.} Figure \ref{fig:ptdlmax} varies the maximum possible side length, $l_{max}$, of uncertainty regions from 1 to 5, where other parameters are set to default values. With larger uncertainty regions, uncertain objects have higher chance of overlapping. Thus, in figures, with the increase of the $l_{max}$ value, both wall clock time and communication cost tend to slightly increase.

\begin{figure}[t!]
\centering%\vspace{-2ex}
\subfigure[][{\small wall clock time}]{\hspace{-4ex}
\scalebox{0.065}[0.065]{\includegraphics{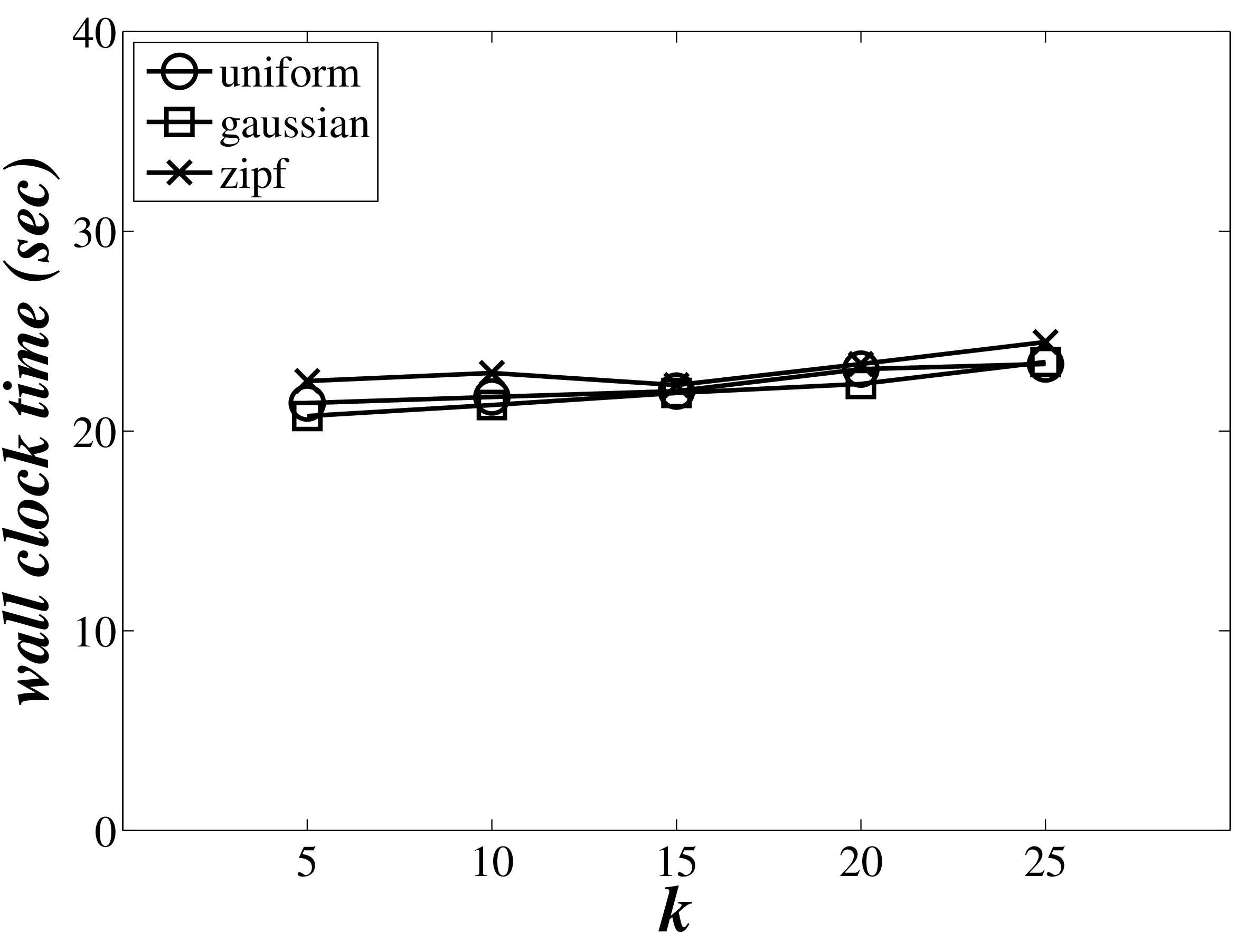}}\label{subfig:ktime}
}\qquad%
\subfigure[][{\small communication cost}]{
\scalebox{0.065}[0.065]{\includegraphics{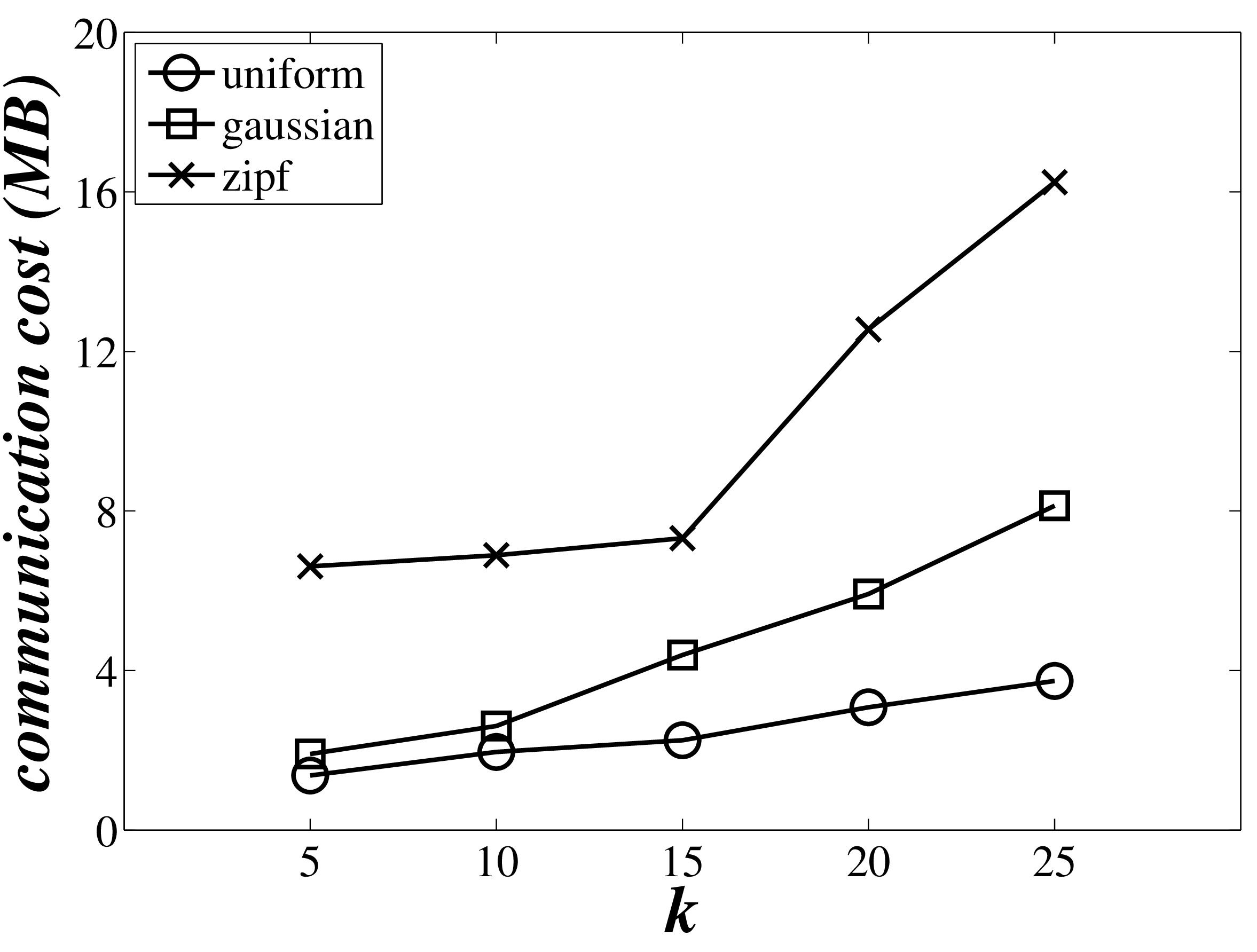}}\label{subfig:kcost}
}%\vspace{-3ex}
\caption{\small The PTD performance vs. parameter $k$.}%\vspace{-2ex}
\label{fig:ptdk}%\vspace{2ex}
\end{figure}

%\vspace{1ex}
\noindent {\bf The PTD performance vs. $k$.} Figure \ref{fig:ptdk} illustrates the effect of parameter $k$ in top-$k$ dominating queries on the performance of our $PTD$ algorithm, where $k = 5, 10, 15, 20$, and $25$ and other parameters are set to default values. When we increase $k$, the wall clock time also smooth increases, whereas the communication cost increases slightly for $uniform$ data set, but sharply for Gaussian and Zipf data distributions (due to the skewed data distributions).

\begin{figure}[t!]
\centering%\vspace{-2ex}
\subfigure[][{\small wall clock time}]{\hspace{-4ex}
\scalebox{0.07}[0.07]{\includegraphics{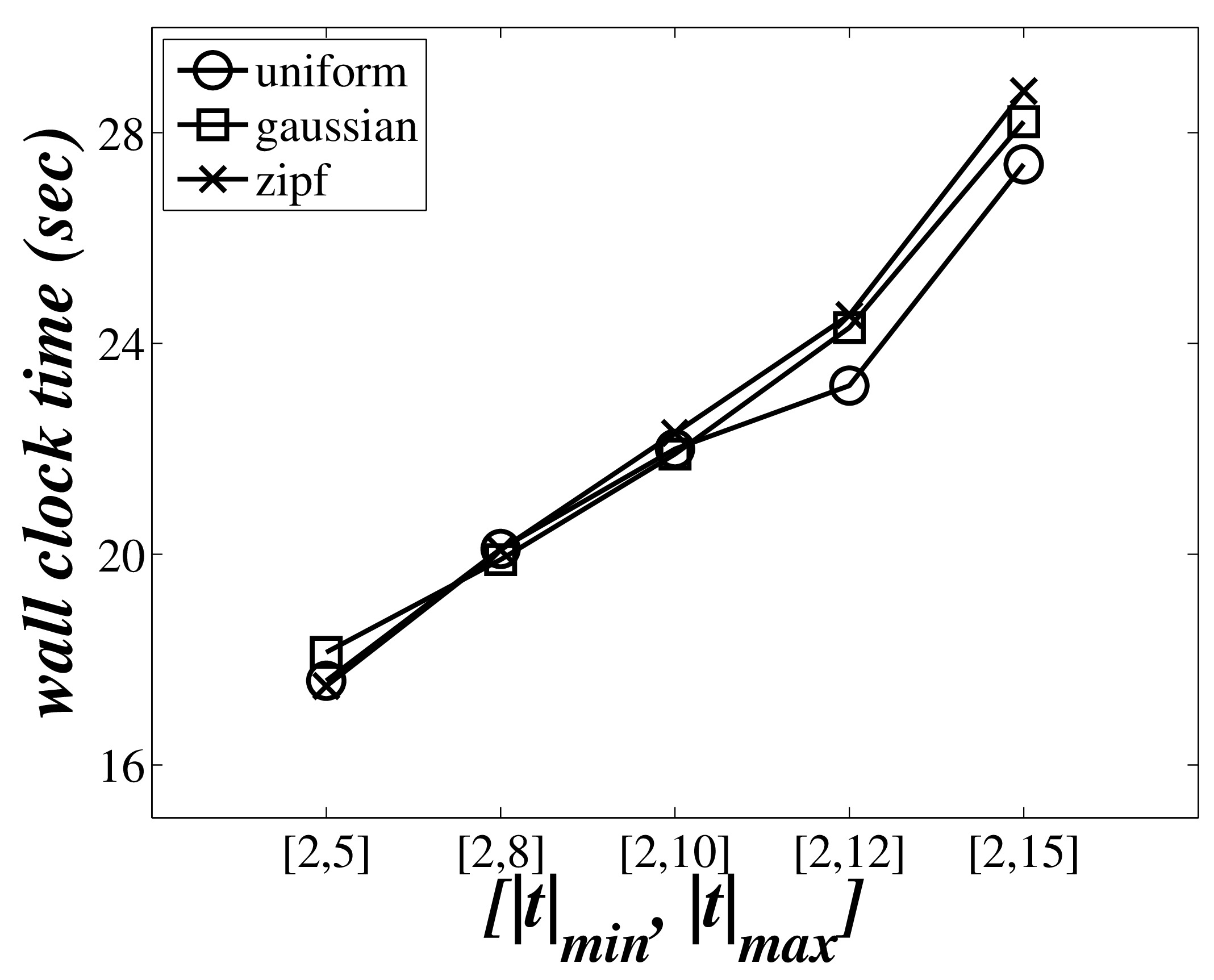}}\label{subfig:dtime}
}\qquad%
\subfigure[][{\small communication cost}]{
\scalebox{0.07}[0.07]{\includegraphics{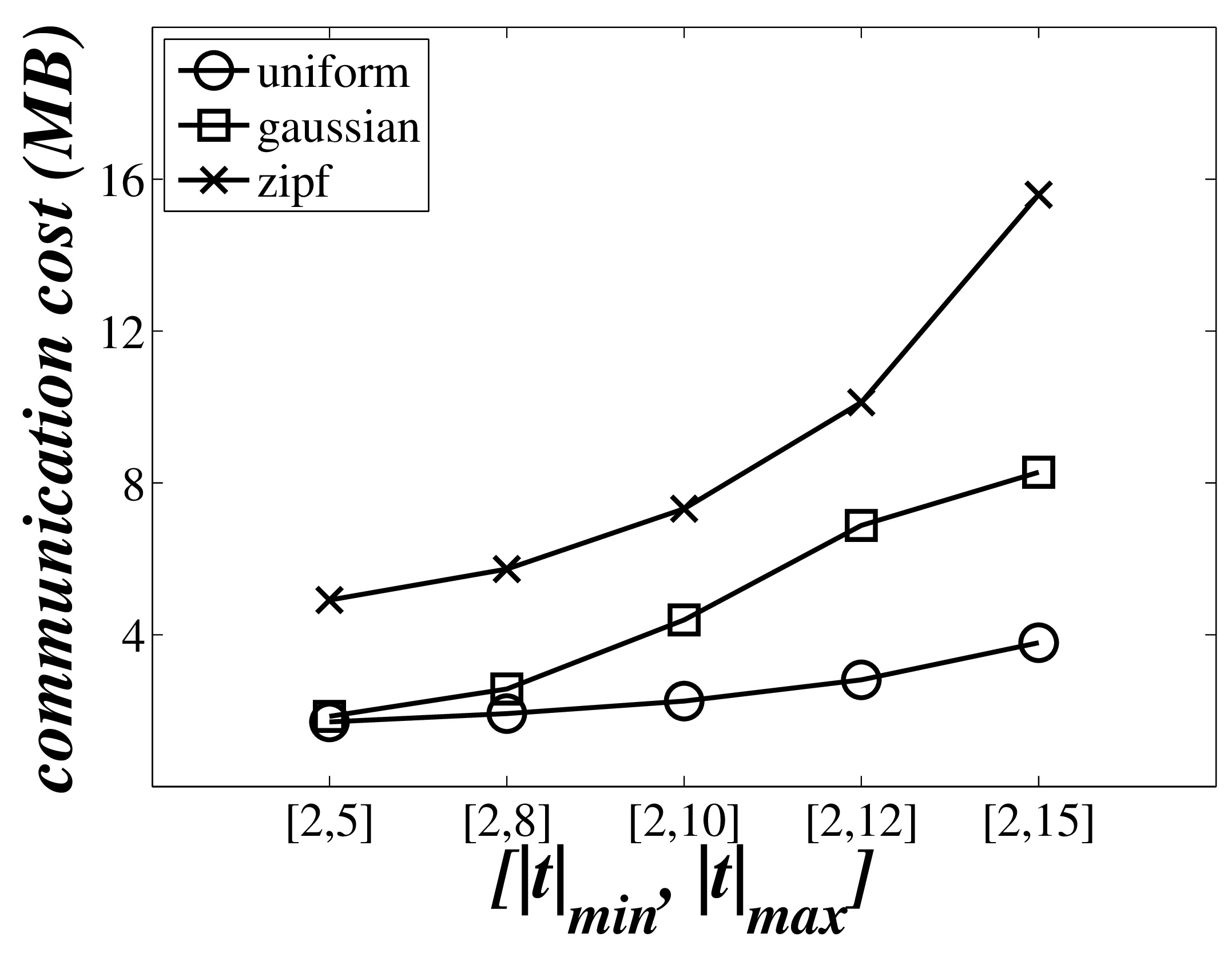}}\label{subfig:dcost}
}%\vspace{-2ex}
\caption{\small The PTD performance vs. the range, $[|t|_{min}, |t|_{max}]$, for No. of instances per uncertain object.}%\vspace{2ex}
\label{fig:ptdinstanc}
\end{figure}

\begin{figure}[t!]
\centering%\vspace{-2ex}
\subfigure[][{\small wall clock time}]{\hspace{-4ex}
\scalebox{0.065}[0.065]{\includegraphics{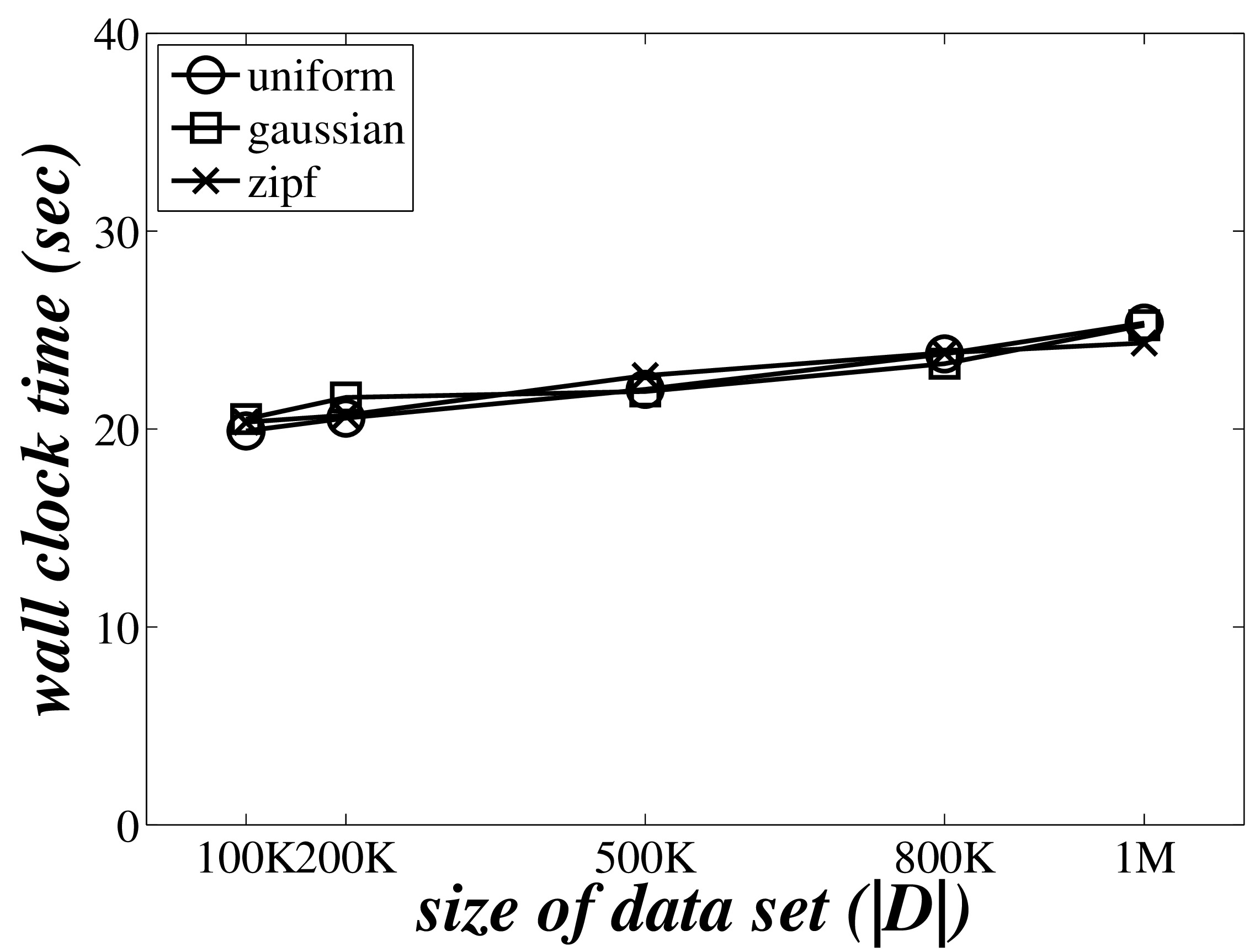}}\label{subfig:ptdtime}
}\qquad%
\subfigure[][{\small communication cost}]{
\scalebox{0.065}[0.065]{\includegraphics{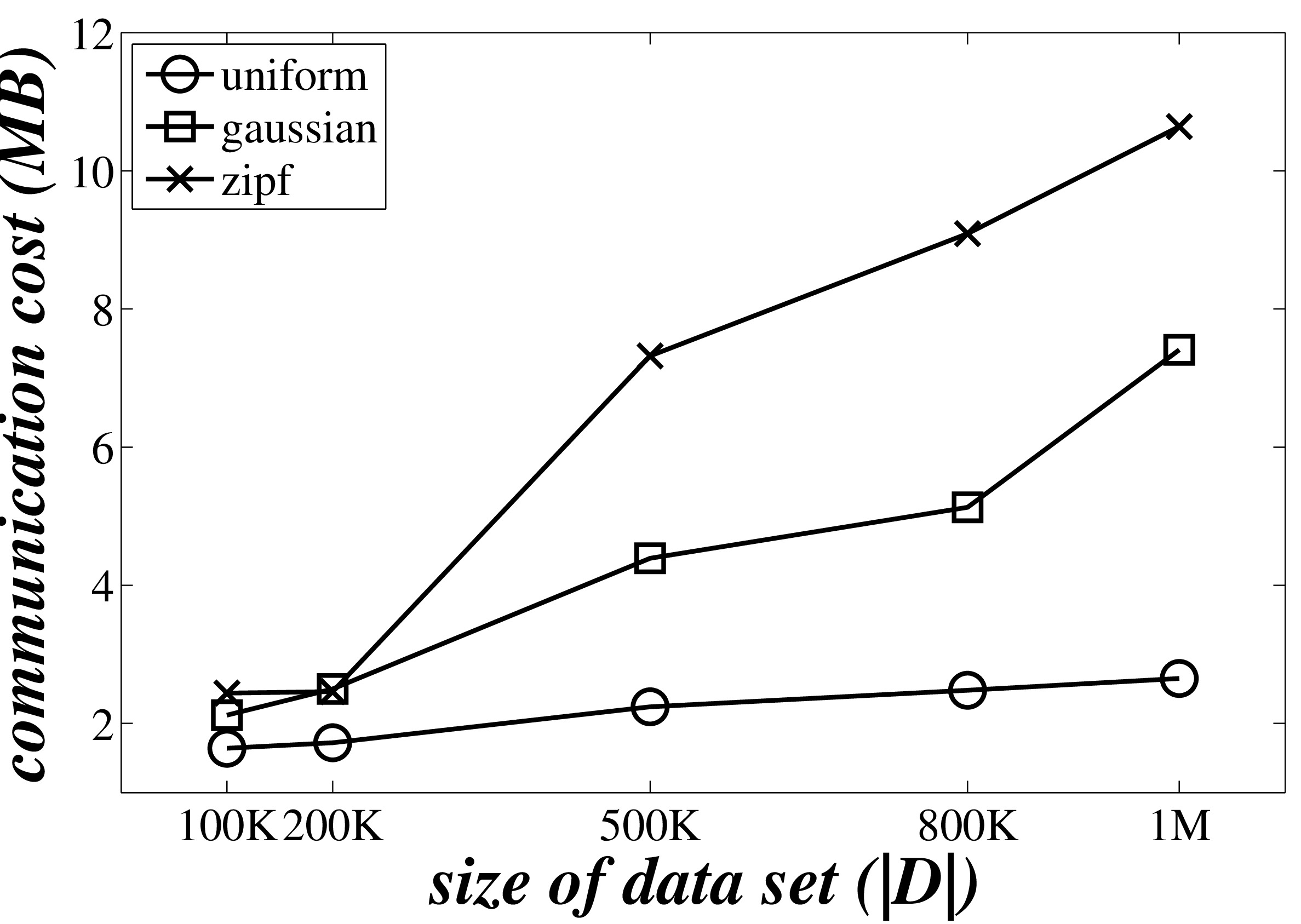}}\label{subfig:ptdcost}
}%\%vspace{-2ex}
\caption{\small The PTD performance vs. the size, $|D|$, of data set.}%\vspace{2ex}
\label{fig:ptddata}
\end{figure}

%\vspace{1ex} 
\noindent {\bf The PTD performance vs. the range, $[|t|_{min}, |t|_{max}]$, for No. of instances per object.} Figure \ref{fig:ptdinstanc} shows the wall clock time and communication cost of our $PTD$ algorithm over synthetic data sets, by varying the range, $[|t|_{min}, |t|_{max}]$, for the number $|t|$ of instances per uncertain object $t$ from $[2, 5]$ to $[2, 15]$, where default values are used for other parameters (as depicted in Table \ref{table:parameter}). The experimental results show that, with more instances per object, the execution time and communication cost are increasing. This is reasonable, since more instances in uncertain objects should be processed and more key-value pairs will be emitted among servers during the distributed PTD query processing. 
%{\color{Xiang} (discuss 1 server)}

%\vspace{1ex} 
\noindent {\bf The PTD performance vs. the number, $|D|$, of uncertain objects.}
Figure \ref{fig:ptddata} evaluates the scalability of our $PTD$ algorithm by varying the number, $|D|$, of uncertain objects from $100K$ to $1M$, where other parameters are set to their default values. Note that, each uncertain object has about 2-10 instances, which indicates that the total number of instances can be up to $2M$ - $10M$ in uncertain database $D$. From figures, when the data size $|D|$ becomes larger, the wall clock time of our $PTD$ algorithm slightly increases (from 20 $sec$ to 25 $sec$), whereas the communication cost increases from about 2 MB to 10 MB. This is reasonable, since larger data sets lead to more candidates and thus higher time/communication costs to filter and refine. 

\begin{figure}[t!]
\centering%\%vspace{-2ex}
\scalebox{0.09}[0.09]{\includegraphics{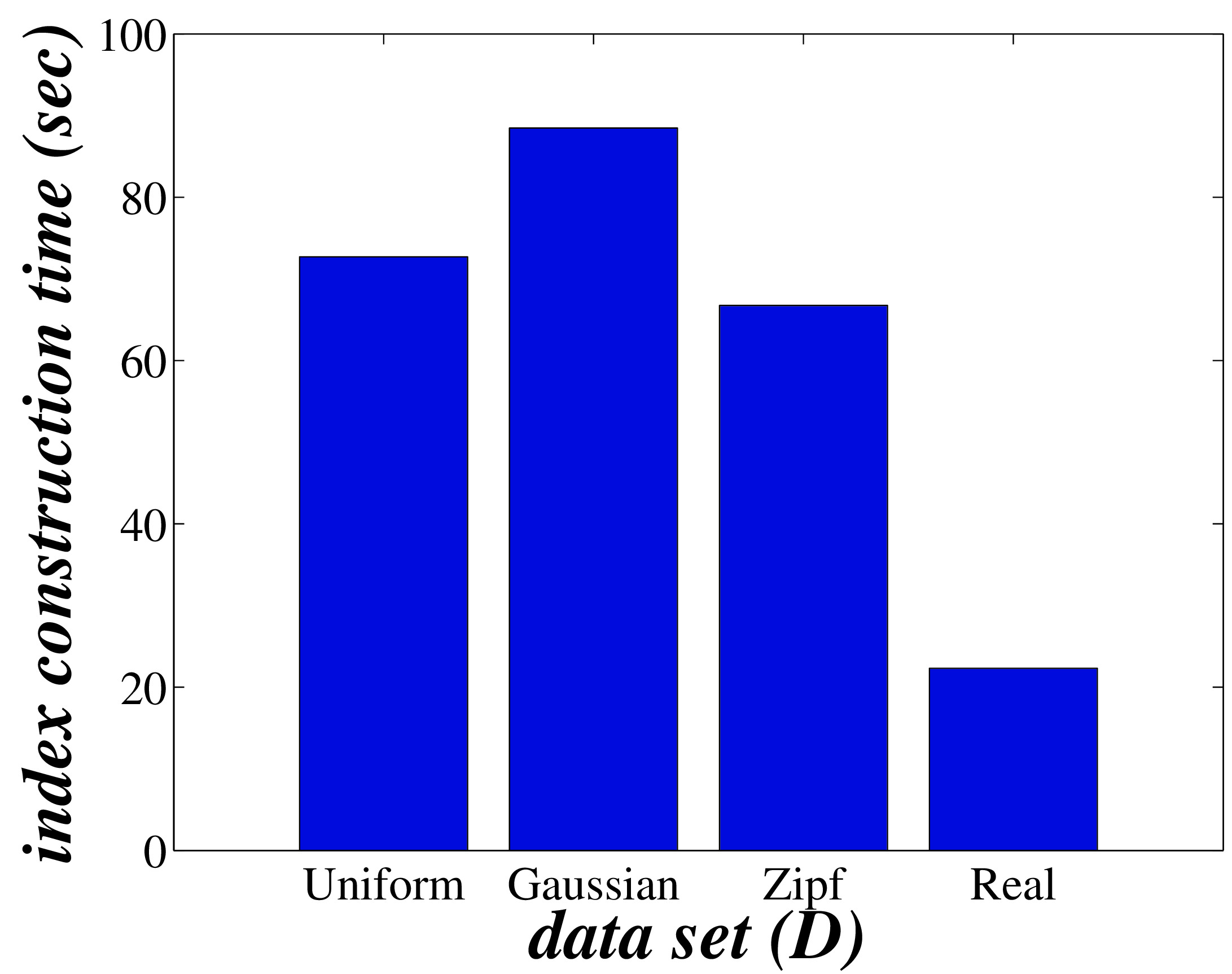}}\label{subfig:indextime}%\%vspace{-1ex}
\caption{\small The index construction time vs. real/synthetic data sets.}%\vspace{2ex}
\label{fig:ptdindexconst}
\end{figure}

%\vspace{1ex} 
\noindent {\bf The index construction time vs. real/synthetic data set $D$.} Finally, we also report the index construction time in Figure \ref{fig:ptdindexconst}, for different real and synthetic data sets, where parameters of synthetic data are set to default values. Note that, the index construction time of the real data set is lower than that of synthetic data, due to smaller size (i.e., $|D| = 98,451$) of real data set (compared with $|D|=500K$ for synthetic data). For all real/synthetic data sets, the offline index construction times are around 22 $\sim$ 88 seconds.

To summarize, our $PTD$ algorithm is scalable and robust against different parameters. We also tested other parameters (e.g., other data distributions) with similar experimental results, which will not be reported here.

%\%vspace{2ex}
\section{Related Work}
\label{sec:related_work}

In this section, we overview previous works on the top-$k$ dominating query in centralized certain/uncertain databases, and query processing in centralized and distributed uncertain databases.

%\vspace{2ex}
\subsection{Top-$k$ Dominating Query on Centralized Certain/Uncertain Data}

%\vspace{1ex}
\noindent {\bf Top-$k$ Dominating Query on Certain Data.} In the centralized environment, Papadias et al. \cite{Papadias05} first proposed a variant of the skyline query, that is, the top-$k$ dominating query, over certain data. The top-$k$ dominating query takes the advantages of both skyline and top-$k$ queries, which is not sensitive to scales of dimensions, does not require users to specify a ranking function, and can control the size $k$ of the returned query answers. Yiu and Mamoulis \cite{yiu} considered the top-$k$ dominating query over certain data, where attributes of objects are static coordinates (which is different from our work that takes into account dynamic attributes of objects w.r.t. query point $q$). In particular, they utilized aggregates in the aR-tree index to estimate the lower/upper bounds of objects' scores, which can be used for pruning and facilitate the top-$k$ dominating query answering. 

In addition, there are some other variants of top-$k$ dominating queries over certain data, such as the top-$k$ dominating query assuming that objects are vertically decomposed \cite{Han15,Tiakas11}, that in the metric space \cite{Tiakas14}, continuous top-$k$ dominating query \cite{Santoso14}, and so on.

%\%vspace{1ex}
\noindent {\bf Top-$k$ Dominating Query on Uncertain Data.} In the literature of uncertain databases, there are some related works \cite{xlian,Zhang10,lian2013probabilistic,Feng13} on probabilistic top-$k$ dominating queries (PTD) under \textit{possible worlds} \cite{dalvi2007efficient} semantics. Specifically, Lian and Chen \cite{xlian,lian2013probabilistic} studied the PTD query (and its variant in arbitrary subspaces) over uncertain objects with dynamic attributes with respect to a query point $q$. Zhang et al. \cite{Zhang10} considered a variant of the PTD query, \textit{threshold-based probabilistic top-$k$ dominating query} (PtopkQ), over uncertain objects with static attributes (i.e., coordinates), which retrieves $k$ objects $o$ with the highest scores, where the score is given by the maximum number of objects dominated by uncertain object $o$ with probability greater than a user-specified threshold. Moreover, Feng et al. \cite{Feng13} studied the PTD query over sliding windows of uncertain data streams.

All prior works mentioned above designed PTD pruning heuristics, and investigated efficient processing of PTD queries via indexes on a single machine. However, in practice, uncertain databases are of large scale in many real-world applications, and/or sometimes uncertain data are collected and stored in distributed servers. Therefore, PTD query processing on centralized uncertain database may not meet the strict requirement of low response time in real applications. Our work in this paper exactly formulates and tackles the PTD problem over uncertain databases in a distributed environment, which cannot directly borrow previous PTD techniques on certain data or centralized uncertain data.

%\%vspace{2ex}
\subsection{Query Processing on Centralized and Distributed Uncertain Databases}

%\vspace{1ex}
\noindent {\bf Centralized Uncertain Data Management.} In the literature, many systems such as MystiQ \cite{Boulos05}, Orion \cite{Cheng05}, TRIO \cite{Benjelloun06}, MayBMS \cite{Antova07},
MCDB \cite{Jampani08}, and BayesStore \cite{Wang08} have been proposed to manipulate probabilistic/uncertain data. The data
uncertainty can be classified into two categories, \cite{Singh08}, \textit{attribute uncertainty} and \textit{tuple uncertainty}. The
attribute uncertainty usually involves (spatial) uncertain databases, whereas the tuple uncertainty is often modeled by
probabilistic (relational) databases. In uncertain databases, each object is represented by a spatial \textit{uncertainty region}, in
which the uncertain object resides following some probabilistic distribution. This object distribution can be captured by either
discrete samples (associated with existence probabilities) \cite{Pei07,Hua08,Lian08} or a continuous \textit{probability
density function} (pdf) \cite{Cheng03,Deshpande04}. In this paper, we consider the tuple uncertainty, that is, probabilistic databases
\cite{dalvi2007efficient}, which consist of $x$-tuples, each containing mutually exclusive tuples (or \textit{alternatives}) with existence
probabilities.

Query processing in uncertain databases has been extensively studied in many real applications. In particular, various probabilistic query types have been explored over uncertain data, including \textit{probabilistic range query} \cite{Cheng03,cheng2004querying}, \textit{probabilistic nearest neighbor query} \cite{Cheng03,cheng2004querying,Kriegel07}, \textit{probabilistic top-$k$ query} \cite{Hua08,Cormode09,Li11}, \textit{probabilistic skyline query} \cite{Pei07}, and so on. In these works, specific for different query types, effective pruning methods are often designed to reduce the problem search space, and uncertain objects are organized by multidimensional indexes to facilitate efficient query processing algorithms on a centralized machine.

Previous techniques on a centralized machine cannot be directly used in our PTD query over distributed servers, since we need to design novel and specific MapReduce functions for the distributed PTD query answering. 

%\vspace{1ex}
\noindent {\bf Distributed Uncertain Data Management.} In the distributed environment, prior works \cite{zhang2016efficient,zhou2016adaptive,park,park2013parallel} studied how to answer skyline and/or reverse skyline queries in parallel over probabilistic data, following the MapReduce framework. Similarly, Li et al. \cite{Li09} tackled the distributed top-$k$ queries over uncertain data. To the best of our knowledge, no previous works tackled the PTD problem over distributed uncertain databases. Due to different problem definitions, we cannot directly apply previous works on distributed (reverse) skyline or top-$k$ queries to solving our PTD problem in a distributed setting.

%\%vspace{2ex}
\section{Conclusion}
\label{sec:conclusion}

In this paper, we study the problem of probabilistic top-$k$ dominating (PTD) query over large-scale and distributed uncertain database. In order to efficiently and effectively tackle the PTD problem, we design a cost model for the index distribution with the purposes of maximizing the pruning power and minimizing the communication cost during the PTD query processing. We also propose an effective MapReduce framework for answering distributed PTD queries, which consists of specifically designed Mapper and Reducer functions that incorporates pruning strategies to reduce the PTD search space. Extensive experiments have been conducted to show the efficiency and effectiveness of our proposed PTD approaches over both real and synthetic data sets.

\section*{Acknowledgement}
\label{sec:acknowledgement}
Funding for this work was provided by NSF OAC No. 1739491 and Lian Startup No. 220981, Kent State University.

\bibliographystyle{spmpsci}      % mathematics and physical sciences
\bibliography{mybibfile}  


\end{document}